\documentclass[]{aastex}

\usepackage{emulateapj5}
\usepackage{onecolfloat}
\usepackage{graphicx} 
\usepackage{fancyheadings} 
\usepackage{ulem}
\usepackage{rotating}
\usepackage{lscape}

\newcommand{\mnhi}{N_{\rm HI}}
\newcommand{\nhi}{$N_{\rm HI}$}

\newcommand{\etal}{et al.\ }

\newcommand{\lya}{Ly$\alpha$}

\newcommand{\kms}{km~s$^{-1}$ }
\newcommand{\cm}[1]{\, {\rm cm^{#1}}}
\newcommand{\N}[1]{{N({\rm #1})}}

\newcommand{\sci}[1]{{\rm \; \times \; 10^{#1}}}
\newcommand{\ltk}{\left [ \,}
\newcommand{\ltp}{\left ( \,}

\newcommand{\rtk}{\, \right  ] }
\newcommand{\rtp}{\, \right  ) }

\newcommand{\ohf}{{1 \over 2}}

\newcommand{\perd}{\;\;\; .}
\newcommand{\cmma}{\;\;\; ,}

\newcommand{\mkms}{{\rm \; km\;s^{-1}}}

\newcommand{\lfeII}{\Lambda_{\rm [FeII]}}

\newcommand{\nfej}[1]{N_{J= #1}({\rm Fe^+})}
\newcommand{\nsij}[1]{N_{J= #1}({\rm Si^+})}
\newcommand{\noij}[1]{N_{J= #1}({\rm O^0})}

\begin{document}

\twocolumn[%
\submitted{Revised Manuscript: April 22, 2006}

\title{Dissecting the Circumstellar Environment of 
$\gamma$-Ray Burst Progenitors}

\author{Jason X. Prochaska\altaffilmark{1},
	Hsiao-Wen Chen\altaffilmark{2},
	and Joshua S. Bloom\altaffilmark{3}}

\begin{abstract} 
We investigate properties of the interstellar medium (ISM)
in galaxies hosting long-duration $\gamma$-ray bursts (GRBs)
from an analysis of atomic species (Mg$^0$, Fe$^0$) and excited
fine-structure levels of ions (e.g.\ Si$^+$).  Our analysis is guided
primarily by echelle observations of GRB~050730 and GRB~051111. These
sightlines exhibit fine-structure transitions of O$^0$,
Si$^+$, and Fe$^+$ gas that have not yet been detected in intervening
quasar absorption line systems.  Our results indicate that the gas
with large \ion{Mg}{1} equivalent width (e.g.\ GRB~051111) must occur
at distances $\gtrsim 50$\,pc from GRB afterglows to avoid
photoionization.  We examine the mechanisms for fine-structure
excitation and find two processes can contribute: (1) indirect UV
pumping by the GRB afterglow provided a far-UV intensity in excess of
10$^6$ times the Galactic radiation field; and (2) collisional
excitation in gas with electron density $n_e > 10^4 \cm{-3}$.  The
observed abundances of excited ions are well explained by UV pumping
with the gas at $r\,\sim$\,a few hundred pc from the afterglow for
GRB~051111 and $r<100$ pc for GRB~050730, without invoking extreme gas
density and temperature in the ISM.  We show that UV pumping alone
provides a simple explanation for all reported detections of excited
ions in GRB afterglow spectra.  
The presence of strong fine-structure transitions therefore
may offer little constraint for the gas density or temperature.
We discuss additional implications of UV pumping including its impact
on chemical abundance measurements, new prospects for observing
line-strength variability, and future prospects for studying the gas
density and temperature.  Finally, we list a series of criteria that
can distinguish between the mechanisms of UV pumping and collisional
excitation.  


\end{abstract}
]

\altaffiltext{1}{UCO/Lick Observatory; University of California, Santa Cruz;
	Santa Cruz, CA 95064; xavier@ucolick.org}
\altaffiltext{2}{Department of Astronomy \& Astrophysics; University of Chicago;
	5640 S.\ Ellis Ave., Chicago, IL 60637; hchen@oddjob.uchicago.edu}
\altaffiltext{3}{Department of Astronomy, 601 Campbell Hall, 
        University of California, Berkeley, CA 94720-3411}

\pagestyle{fancyplain}
\lhead[\fancyplain{}{\thepage}]{\fancyplain{}{PROCHASKA, CHEN, \& BLOOM}}
\rhead[\fancyplain{}{Unique Extragalactic Sightlines}]{\fancyplain{}{\thepage}}
\setlength{\headrulewidth=0pt}
\cfoot{}

\section{Introduction}

Optical afterglows of long-duration $\gamma$-ray bursts (GRBs) are
almost certainly signposts of extreme star-forming regions at high
redshift, because these bursts originate in the catastrophic death of
massive stars \citep[e.g.][]{woo93,pac98a,bkp+02,smg+03}.  A detailed
study of metal absorption features identified in the circumburst
environment can, in principle, yield strong constraints on the
progenitor \citep[e.g.][]{pl98a,rgs+05,vlg05} and offer important
insights for understanding mass-loss and chemical feedback during the
final evolution stages of massive stars \citep{wij01,vink05}. The
benefit of GRBs, which presumably are embedded in environments of
Wolf-Rayet stars, is that their optical afterglows provide extremely
bright lighthouses at moderate to high redshift, allowing redshifted
UV lines to be (albeit transiently) observable with high-resolution
spectroscopy.

Past studies based on low-resolution afterglow spectra have been
limited to a few diagnostic measurements such as the \ion{H}{1} column
density, the gas metallicity, and the dust-to-gas ratio
\citep[e.g.][]{sff03,vel+04}.  In contrast, high-resolution
spectroscopy of GRB afterglows has uncovered detailed kinematic
signatures, population ratios of excited ions, and chemical
compositions of the interstellar medium (ISM) and, perhaps, the
circumstellar medium (CSM) of the progenitor star of the GRB
\citep{vel+04,cpb+05}.  These quantities, in principle, permit direct
comparisons with simulations describing the evolution history of GRB
progenitors \citep[e.g.][]{vlg05}.

An emerging feature of GRB progenitor environments is the presence of
strong fine-structure transitions from excited states of C$^+$,
Si$^+$, O$^0$, and Fe$^+$ \citep{cpb+05}.  Together with their
resonance transitions, these fine-structure lines can in principle
provide robust constraints on the temperature and density of the gas,
as well as the ambient radiation field, independent of the ionization
fraction, relative abundances, or metallicity of the gas \citep{bw68}.
In particular, detections of \ion{Fe}{2} fine structure transitions
have only been reported in rare places such as broad absorption-line
(BAL) quasars \citep{has+02}, $\eta$ Carinae \citep{gull05,nielsen05}
and the circumstellar disk of $\beta$ Pictoris \citep{lvf88}.  The
presence of strong \ion{Fe}{2} fine-structure lines therefore suggests
extreme gas density and temperature in the GRB progenitor environment.
In the case of circumburst medium, however, the greatly enhanced
radiation field due to the GRB afterglow may also contribute
significantly to the formation of these excited ions.

\begin{figure*}
\begin{center}
\includegraphics[width=6.5in]{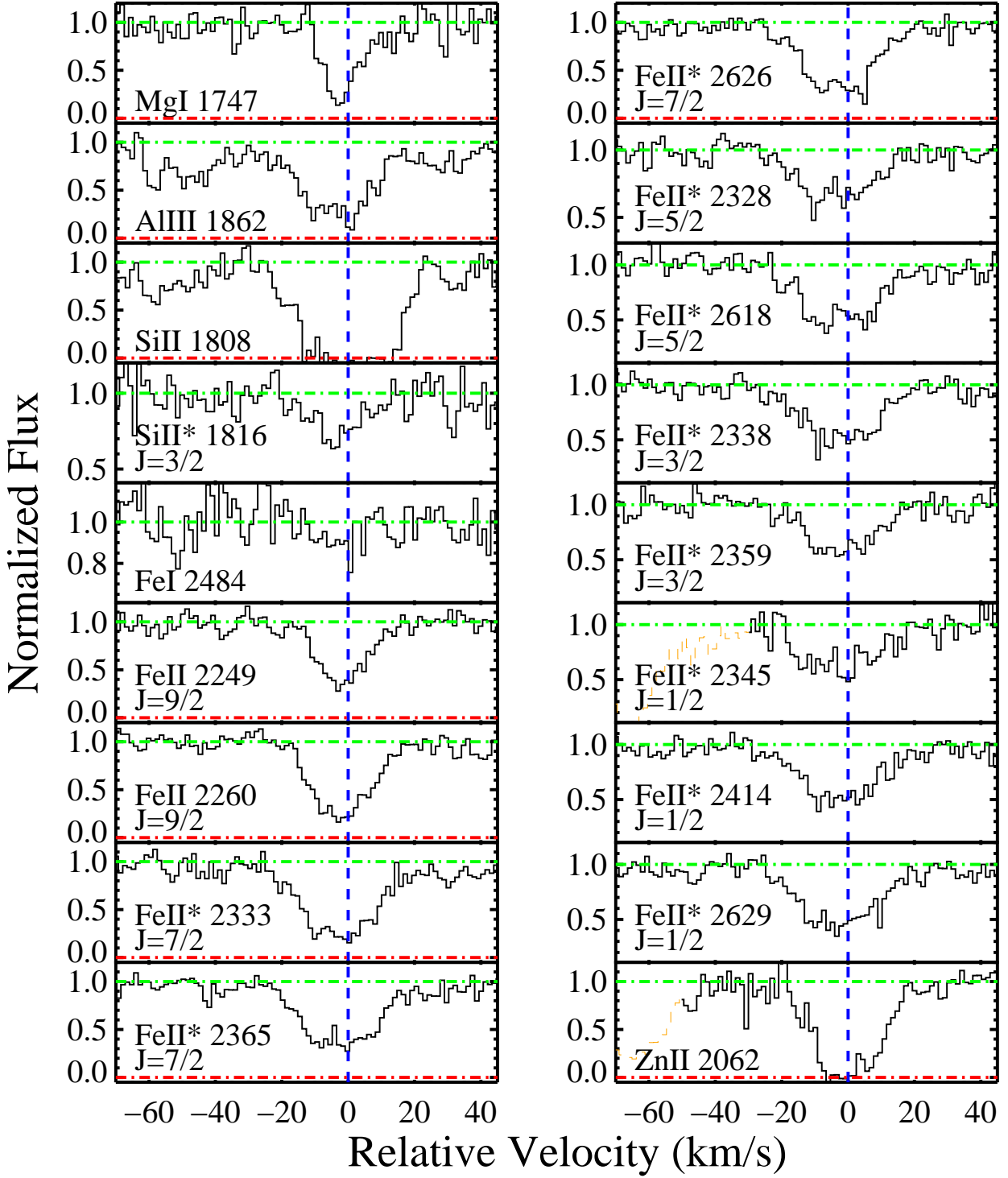}
\caption{Velocity profiles for transitions arising from 
gas in the host galaxy of GRB~051111.  The vertical dashed line at $v=0$
corresponds to $z=1.5495$ and the dotted lines indicate blends
with coincident transitions.
}
\label{fig:051111}
\end{center}
\end{figure*}

In this paper we present a comprehensive study of the physical
properties of the ISM in GRB host galaxies, using absorption lines
identified in the echelle spectra of two GRB afterglows, GRB\,051111
and GRB\,050730 as case studies.  We first consider the enhanced
radiation field in the circumburst environment due to the progenitor
star as well as the optical afterglow, and constrain the distance of
the observed neutral gas cloud based on the presence/absence of
various atomic lines such as Mg\,I and Fe\,I.  The dominant species of
these elements is in the singly ionized state, because their first
ionization potential is less than 1\,Ryd.  The presence of strong
atomic lines therefore places a lower limit to the distance of the
cloud from the afterglow.  Next, we investigate the excitation
mechanism of the absorbing gas, again taking into account the presence
of an intensified radiation field known from afterglow light-curve
observations.  We compare the predicted population ratios with
observations of different ions under photon pumping and collisional
excitation scenarios.  Although the analysis is based on echelle data
obtained for two GRB hosts, we note that the results are applicable to
the majority of long-duration GRB's discovered to date.  Finally, we
discuss additional framework for studying the ISM gas of future
events.

The paper is organized as follows.  The observations and analysis
of echelle spectra for GRB~051111 and GRB~050730 are briefly summarized
in $\S$~\ref{sec:obs}.  In $\S$~\ref{sec:rad} we describe the
radiation field produced by GRB afterglows.  In $\S$~\ref{sec:neut}
we analyze the non-dominant atomic transitions observed in optical
afterglow spectra, while $\S$~\ref{sec:fine} investigates the
mechanisms for populating fine-structure states in the dominant ions.
In $\S$~\ref{sec:disting} we consider tests to distinguish
between UV pumping and collisional excitation as the principal 
mechanism for population of fine-structure levels.  Finally, 
$\S$~\ref{sec:discuss} presents a discussion of the results
for the specific cases of GRB~051111 and GRB~050730 and, also,
generic results for observations of long-duration GRBs.
Throughout the paper we assume a cosmology with 
$\Omega_\Lambda = 0.7$, $\Omega_m = 0.3$ and 
$H_0 = 70 {\rm km \, s^{-1} \, Mpc^{-1}}$.

\begin{figure}[ht]
\begin{center}
\includegraphics[width=3.5in]{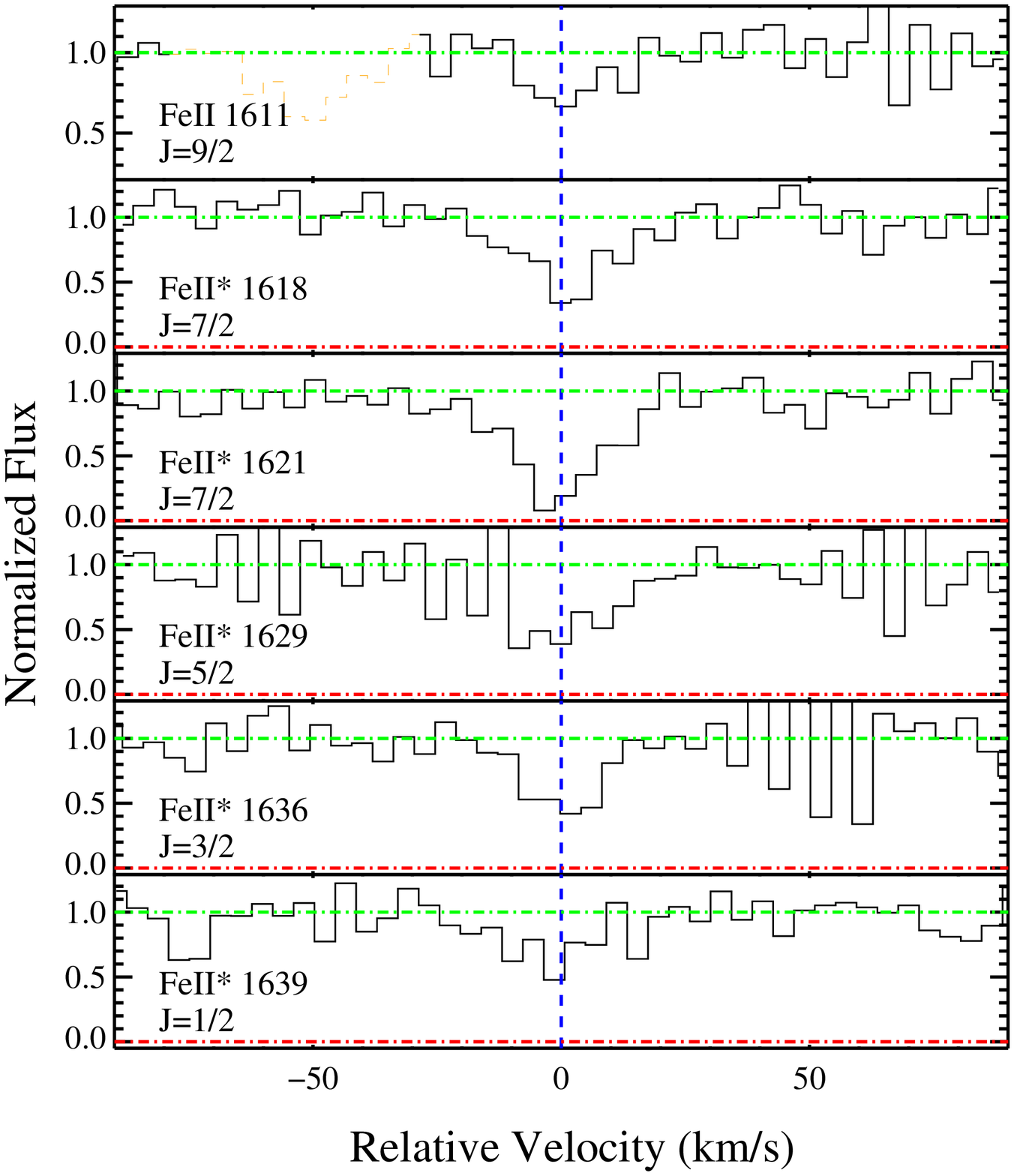}
\caption{Velocity profiles for \ion{Fe}{2} transitions arising from 
gas in the host galaxy of GRB~050730.  The vertical dashed line at $v=0$
corresponds to $z=3.96855$ and the dotted lines indicate blends
with coincident transitions.
}
\label{fig:050730}
\end{center}
\end{figure}

\section{Observations and Column Densities}
\label{sec:obs}

Echelle spectroscopic observations of GRB\,051111 were obtained using
the High Resolution Echelle Spectrometer \citep[HIRES;][]{vogt94} on
the Keck~I telescope.  The observations consisted of three exposures
of 1800 s duration each and were initiated by the Keck Observatory
staff at UT 07:02:57.92 on 11 November 2005.  The spectra\footnote{The
data are publicly available at http://www.graasp.org} covered a
wavelength range from $\lambda = 4165 - 8720$\AA\ with a spectral FWHM
resolution of $\approx 5 \mkms$.  A summary of the data reduction is
presented in a separate paper (Prochaska \etal\ 2006, in preparation).
The final coadded spectrum was normalized to a unit continuum and has
${\rm SNR \approx 15}$ per 1.3~\kms\ pixel.  Figure~\ref{fig:051111}
presents the subset of transitions related to the GRB for the analysis
performed in this paper.  We \citep{gcn4255} reported the discovery of
the redshift $z=1.549$ for GRB\,051111 and preliminary analysis of the
results \cite{gcn4271}.  %\cite{berger051111} and 
\cite{penprase05} present an independent analysis of the same data.

Echelle spectroscopic observations of GRB\,050730 were obtained using
the MIKE echelle spectrograph \citep{bernstein03} on the Magellan Clay
telescope.  The data cover observed wavelengths $\lambda = 3400 -
9000$\AA\ at a FWHM~$\approx 11$\kms\ dispersion with a typical SNR=12
per resolution element.  A detailed description of the observation and
data reduction is presented in \cite{cpb+05}.  In addition to the
absorption features presented in Chen et al., here we also present
identifications of \ion{Fe}{2} transitions from excited fine structure
states, $3d^6\,4s\;^6D_{J=5/2,3/2,1/2}$ (Figure~\ref{fig:050730}).
Column density measurements of the two GRB sightlines are presented in
Tables~\ref{tab:051111} and \ref{tab:050730}.  The atomic data are
taken from
\cite{morton91,bergeson93,bml94,bergeson96,tripp96,raassen98,morton03}.

\begin{table*}[ht]\footnotesize
\begin{center}
\caption{{\sc IONIC COLUMN DENSITIES FOR GRB\,051111\label{tab:051111}}}
\begin{tabular}{lccccccc}
\tableline
\tableline
Ion & J$^a$ & $E_{low}$ & $\lambda$ & $f$
& $\log N$ & $\log N_{adopt}$ \\
& & cm$^{-1}$ & (\AA) & &  \\
\tableline
\ion{Mg}{1}\\
& &    0.00 & 1747.794 &    0.009080 &$14.68 \pm 0.04$&$ 14.68 \pm 0.04$\\
& &    0.00 & 1827.935 &    0.023900 &$>14.63$&\\
& &    0.00 & 2026.477 &    0.112000 &$>14.31$&\\
\ion{Al}{3}\\
& &    0.00 & 1862.790 &    0.268000 &$13.51 \pm 0.02$&$ 13.51 \pm 0.02$\\
\ion{Si}{2}\\
&1/2 &    0.00 & 1808.013 &    0.002186 &$>16.07$&$> 16.07$\\
&1/2 &    0.00 & 2335.123 &    0.000004 &$<17.27$&\\
&3/2 &  287.24 & 1816.928 &    0.001660 &$15.00 \pm 0.06$&$ 15.00 \pm 0.06$\\
\ion{Fe}{1}\\
& &    0.00 & 2484.021 &    0.557000 &$<11.75$&$< 11.75$\\
\ion{Fe}{2}\\
&9/2 &    0.00 & 2249.877 &    0.001821 &$15.20 \pm 0.02$&$ 15.24 \pm 0.01$\\
&9/2 &    0.00 & 2260.780 &    0.002440 &$15.28 \pm 0.02$&\\
&7/2 &  384.79 & 2333.516 &    0.069170 &$13.97 \pm 0.01$&$ 13.98 \pm 0.01$\\
&7/2 &  384.79 & 2365.552 &    0.049500 &$13.99 \pm 0.01$&\\
&7/2 &  384.79 & 2383.788 &    0.005175 &$14.05 \pm 0.09$&\\
&7/2 &  384.79 & 2389.358 &    0.082500 &$14.01 \pm 0.02$&\\
&7/2 &  384.79 & 2626.451 &    0.044100 &$<14.00$&\\
&5/2 &  667.68 & 2328.111 &    0.035550 &$13.75 \pm 0.02$&$ 13.76 \pm 0.01$\\
&5/2 &  667.68 & 2607.866 &    0.118000 &$13.76 \pm 0.02$&\\
&5/2 &  667.68 & 2618.399 &    0.050500 &$13.76 \pm 0.02$&\\
&3/2 &  862.61 & 2338.725 &    0.092500 &$13.54 \pm 0.02$&$ 13.55 \pm 0.01$\\
&3/2 &  862.61 & 2359.828 &    0.057270 &$13.59 \pm 0.02$&\\
&3/2 &  862.61 & 2621.191 &    0.003920 &$<13.86$&\\
&1/2 &  977.05 & 2345.001 &    0.157500 &$13.17 \pm 0.03$&$ 13.25 \pm 0.01$\\
&1/2 &  977.05 & 2414.045 &    0.175500 &$13.25 \pm 0.02$&\\
&1/2 &  977.05 & 2622.452 &    0.056000 &$13.28 \pm 0.04$&\\
&1/2 &  977.05 & 2629.078 &    0.173000 &$13.28 \pm 0.02$&\\
\ion{Zn}{2}\\
& &    0.00 & 2026.136 &    0.489000 &$>13.60$&$> 13.71$\\
& &    0.00 & 2062.664 &    0.256000 &$>13.71$&\\
\tableline
\end{tabular}
\end{center}
\tablenotetext{a}{J value for ions with excited states.
$E_{low}$ is the energy above the ground \\ state.}
\tablenotetext{b}{Although the statistical uncertainty is
 as low as 0.01~dex in several cases, \\ we adopt a minimum 
uncertainty of $5\%$ in the analysis.}
\end{table*}

\begin{table*}[ht]\footnotesize
\begin{center}
\caption{{\sc IONIC COLUMN DENSITIES FOR GRB\,050730\label{tab:050730}}}
\begin{tabular}{lccccccc}
\tableline
\tableline
Ion & J$^a$ & $E_{low}$ & $\lambda$ & $f$
& $\log N$ & $\log N_{adopt}$ \\
& & cm$^{-1}$ & (\AA) & &  \\
\tableline
\ion{C}{1}\\
&1/2 &    0.00 & 1277.245 &    0.096650 &$<13.37$&$< 13.37$\\
&1/2 &    0.00 & 1656.928 &    0.140500 &$<13.47$&\\
\ion{O}{1}\\
&1/2 &    0.00 & 1302.168 &    0.048870 &$>15.27$&$> 15.27$\\
&1/2 &    0.00 & 1355.598 &    0.000001 &$<18.09$&\\
&2/2 &  158.26 & 1304.858 &    0.048770 &$>14.94$&$> 14.94$\\
&4/2 &  226.98 & 1306.029 &    0.048730 &$>14.61$&$> 14.61$\\
\ion{Si}{2}\\
&1/2 &    0.00 & 1260.422 &    1.007000 &$>14.08$&$> 14.69$\\
&1/2 &    0.00 & 1304.370 &    0.094000 &$>14.69$&\\
&1/2 &    0.00 & 1526.707 &    0.127000 &$>14.55$&\\
&1/2 &    0.00 & 1808.013 &    0.002186 &$<15.59$&\\
&3/2 &  287.24 & 1264.738 &    0.903400 &$>14.03$&$< 16.03$\\
&3/2 &  287.24 & 1265.002 &    0.100400 &$>14.56$&\\
&3/2 &  287.24 & 1309.276 &    0.146800 &$>14.28$&\\
&3/2 &  287.24 & 1817.451 &    0.000129 &$<16.03$&\\
\ion{Fe}{2}\\
&9/2 &    0.00 & 1608.451 &    0.058000 &$>14.72$&$ 14.98 \pm 0.12$\\
&9/2 &    0.00 & 1611.200 &    0.001360 &$14.98 \pm 0.12$&\\
&7/2 &  384.79 & 1618.468 &    0.021400 &$14.26 \pm 0.06$&$ 14.29 \pm 0.04$\\
&7/2 &  384.79 & 1621.686 &    0.038100 &$14.34 \pm 0.07$&\\
&5/2 &  667.68 & 1629.160 &    0.036700 &$14.12 \pm 0.07$&$ 14.12 \pm 0.07$\\
&3/2 &  862.61 & 1634.350 &    0.020500 &$13.95 \pm 0.11$&$ 13.92 \pm 0.06$\\
&3/2 &  862.61 & 1636.331 &    0.040500 &$13.91 \pm 0.07$&\\
&1/2 &  977.05 & 1639.401 &    0.057900 &$13.65 \pm 0.07$&$ 13.65 \pm 0.07$\\
\tableline
\end{tabular}
\end{center}
\tablenotetext{a}{J value for ions with excited states.
$E_{low}$ is the energy above the ground \\ state.}
\tablenotetext{b}{Although the statistical uncertainty is
 as low as 0.01~dex in several cases, \\ we adopt a minimum 
uncertainty of $5\%$ in the analysis.}
\end{table*}

Column densities of the unblended transitions were measured using the
apparent optical depth method \citep[AODM;][]{savage91,jenkins96} over
a velocity interval $\pm 40 \mkms$ centered at $z=3.96855$ for
GRB\,050730 and at $z=1.54948$ for GRB\,051111.  In Tables 1 \& 2, we
list the column densities of individual transitions, the $1\sigma$
statistical uncertainty, and the weighted mean.  In addition, we
report conservative lower limits for saturated transitions and
$3\sigma$ upper limits for non-detections.  No systematic trend
between column density and oscillator strength is identified in the
measurements, indicating that the absorption lines are well resolved
and that the column density measurements are robust.  We note,
however, that uncertainties in the continuum level are large for
features with rest-frame absorption equivalent width $< 10$ m\AA.  In
spite of a formal statistical error $< 0.01$\,dex in several cases, we
adopt a minimum $1\sigma$ error of $5\%$ in subsequent analyses.

\section{Radiation Field Produced by the GRB Afterglow}
\label{sec:rad}

The intimate connection between long-duration GRBs and the end stage
of the lifetime of massive stars implies that the gaseous clouds in
the vicinity of the progenitor stars have been blasted with intense
ultraviolet radiation fields, especially in the early afterglow phase.
The radiation field near the GRB, significantly enhanced with respect
to the ambient radiation intensity of the host galaxy, modifies the
ionization and excitation states of the gas.  It is therefore critical
to consider first the radiation field produced by the GRB afterglow
before then analyzing the absorption-line profiles.

We adopt the afterglow of GRB~051111 as an example.  \cite{butler06}
have analyzed the light curve data of this event and characterize the
flux of the afterglow at Earth for $t_{\rm obs}>50$ s as
\begin{equation}
F_\nu^{\rm obs} = 1.4 \sci{-25} \ltp \frac{t_{\rm obs}}{50\,\rm s} \rtp^{-\alpha_t}
\ltp \frac{\lambda_{\rm obs}}{6588\,{\rm \AA}} \rtp^{-\beta}  
{\rm erg \, cm^{-2} \, s^{-1} \, Hz^{-1}},
\label{eqn:flux}
\end{equation}
where $\lambda_{\rm obs}$ and $t_{\rm obs}$ are the observed
wavelength and time at Earth.  
The fading of the afterglow is best-fit
with $\alpha_t=0.87$.  
The spectral index was estimated to be
$\beta=0.6$ from a fit to the VRIJHK bands, i.e.\ $h\nu < 6$eV at the
rest frame of the GRB host.  
If we assume that this power-law extends to
higher energies, then Equation~\ref{eqn:flux} describes the
unattenuated flux to energies of several Ryd as a function of time.
At some level this radiation field is attenuated by (1) dust and gas
in the host galaxy: Lyman limit absorption at $h\nu >$~1Ryd, (2) dust
extinction at all wavelengths, and (3) line opacity by heavy elements
and hydrogen Lyman series (concentrated at energies $h\nu > 6$eV).
All of these sources of opacity are expected to contribute and we will
discuss them when relevant.

We consider three energy intervals important to UV/optical absorption
line spectroscopy: (1) a 1\AA\ interval centered at
$h\nu = 0.048$\,eV ($385 \cm{-1}$); (2) 8\,eV
$< h\nu < 13.6$\,eV; and (3) 1\,Ryd $< h\nu < $2\,Ryd.  The first is
set by the energy necessary to populate the first excited state
$J=7/2$ of Fe$^+$ from its ground state $J=9/2$ via a magnetic dipole
transition.  The 1\AA\ interval is chosen to roughly coincide with the 
width of the line-profile which is set by the gas kinematics along
the GRB sightline.  The second interval
is a measure of the far-UV radiation field.  To provide comparison with
the Milky Way, we adopt Habing's constant $G_0 = 1.6 \sci{-3} \, {\rm
erg \, cm^{-2} \, s^{-1}} $, which roughly corresponds to the Galactic
far-UV intensity \citep{habing68}.  More recent estimates place the Galactic
radiation field at $\approx 1.7 G_0$ \citep{gpw80}.  
%The intensity $G_0$
%corresponds to a photon density $n_\gamma \approx 0.005 \cm{-3}$.  
The third energy interval characterizes the radiation field responsible
for ionizing hydrogen atoms.

\begin{figure}[ht]
\begin{center}
\includegraphics[height=3.6in,angle=90]{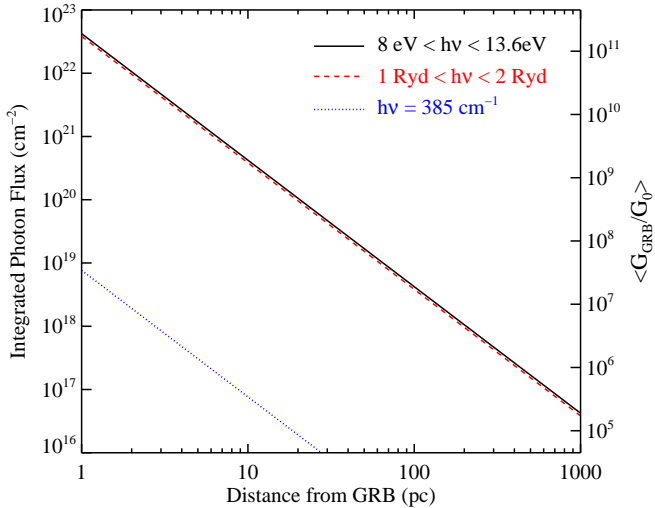}
\caption{Integrated, unattenuated flux of 
photons emitted by the afterglow of GRB~051111
from $t=19.6$ to 1506s with energies in three intervals:
(1) Hydrogen ionizing radiation (1Ryd~$< h\nu < $2Ryd; red curve);
(2) far-UV radiation (8eV~$< h\nu < $13.6eV; black curve);
(3) photons which would excite Fe$^+$ from the $J=9/2$ to $J=7/2$
level (a 1\AA\ interval centered at $h\nu = 0.048$eV; blue curve).
The radial dependence is simple $r^2$ dimming.
The right axis presents the time-averaged intensity of the far-UV
radiation field $G_{GRB}$ relative to Habing's constant $G_0$.
}
\end{center}
\label{fig:radiation}
\end{figure}

In Figure~\ref{fig:radiation} we plot the number of photons per cm$^2$
emitted by the afterglow of GRB~051111 from $t=t_{\rm obs}/(1+z)=19$s 
to 1506s versus
the radial distance $r$ from the GRB in the rest-frame of the GRB host.
%(see Appendix~\ref{app:cosm}).  
For the $E = 385 \cm{-1}$ photons, we
assume a 1\AA\ interval.  %Again, these curves assume no attenuation of
%the radiation field described by Equation~\ref{eqn:flux}.  
Similar to
previous work \citep[e.g.][]{pl98a,draine02}, we find that in the
absence of intervening gas and dust the GRB afterglow would 
initiate a burning front that would ionize an \ion{H}{1}
column of several 10$^{22} \cm{-2}$ atoms at $r=1$\,pc.  As
impressive, one notes that the far-UV intensity is over 10$^5$ times
larger than the ambient Milky Way field even at $r=1$\,kpc.  Despite
its transient nature, the afterglow emits more UV photons in a few
minutes than the entire galaxy in one year (assuming a star formation
rate SFR = 20\,M$_\odot$ yr$^{-1} = 1.6 \sci{28} {\rm erg \; s^{-1} \;
Hz^{-1}}$; e.g.\ Kennicutt 1998).

The estimated strength of the radiation field due to this afterglow has
three important implications, which are applicable to many other GRBs.
First, if neutral gas were present at $r<1$\,pc of the progenitor
prior to the burst (which is unlikely assuming significant UV
luminosity from the GRB progenitor), the implied mean hydrogen density
$<n_H>$ would have to exceed $10^4 \cm{-3}$ for any neutral gas to
survive the burst.  At $r>10$\,pc, however, significant absorption is
still expected from neutral gas if $<n_H>$ is greater than $10\cm{-3}$.
Second, neutral atoms with ionization potential less than 1\,Ryd
(present within the spectrum of GRB~051111) are very unlikely to
survive the first few hundred seconds ($\S$~\ref{sec:neut}).  Third,
the intensity of the radiation field at even 1\,kpc is large enough
that photon pumping of fine-structure states cannot be ignored.
%It is worth noting that the afterglow of GRB~051111 is not
%extraordinary.  Its intrinsic brightness is characteristic of the GRB
%afterglows for which early-time spectroscopic data are available.
%Therefore, the implications of an enhanced, transient UV radiation field apply to
%many other GRB.  
In the following sections, we will consider photon pumping
and discuss constraints on the distance to the progenitor stars, the 
gas density and temperature of the observed gas.

\section{Gaseous Properties As Constrained by Atomic Species}
\label{sec:neut}

By definition, neutral hydrogen gas clouds are opaque to radiation
with $h\nu > 1$\,Ryd and transparent to lower energy photons (aside
from opacity by the Lyman series).  Because the first ionization
potential of most elements is less than 1\,Ryd, the dominant species
of these elements is singly ionized gas, e.g.\ Fe$^+$, Si$^+$, and
Mg$^+$.  However, the atomic phase is detected in trace quantities in
various gaseous phases of the ISM of nearby galaxies
\citep[e.g.][]{frisch90,welty03}.  One of the most common examples is
Mg$^0$.  Surveys of \ion{Mg}{1} lines in the Milky Way have revealed
this atom in both cold and warm phases of the neutral ISM
\citep[e.g.][]{frisch87,frisch90}.  In addition, atomic Mg is also
often detected in quasar absorption line systems exhibiting strong
\ion{Mg}{2} absorption \citep[e.g.][]{steidel92,churchill00a}.  This
includes damped \lya\ systems at all redshifts \citep{rao00,pw02}.
The frequent presence of Mg$^0$ is driven by a relatively
low photoionization
cross-section and modest recombination coefficients.  At low
temperature radiative recombination dominates while at $T \gtrsim
5000$K dielectronic recombination becomes significant, allowing
\ion{Mg}{1} absorption to be observed from even partially ionized
regions \citep{york79}.

In this section, we examine the presence/absence of Mg$^0$ (IP=7.64eV)
and Fe$^0$ (IP=7.90eV) along the GRB~051111 sightline.  Although the
results are directed to the gas in this GRB host galaxy, neutral
species have been reported in other GRB afterglow spectra
\citep[e.g.][]{bsc+03,foley06} and our conclusions are qualitatively
applicable to those cases.  One can also generalize the results to
additional atomic species, e.g., Si$^0$ and C$^0$.

Column density measurements for the atomic species shown in
Figure~\ref{fig:051111} are presented in Table~\ref{tab:051111}.  We
detect several transitions of \ion{Mg}{1} but no \ion{Fe}{1}
transition.  The Mg$^0$ column density for the host galaxy of
GRB~051111 may be the largest ever recorded outside the Milky Way.
%  Undoubtedly, this is at least partly due to the very large
%column densities of the dominant ionic species as evidenced by the
%saturated \ion{Zn}{2} $\lambda\lambda 2026, 2062$ and \ion{Si}{2}
%$\lambda 1808$ profiles (Figure~\ref{fig:051111}).  
Based on the kinematics of the \ion{Mg}{1} profile and its large
column density, we argue that this atomic gas is co-spatial with the
majority of low-ion absorption observed along the sightline.  As such,
we will treat the Mg$^0$, Mg$^+$, Fe$^0$, and Fe$^+$ gas to arise from
a single cloud within the ISM of the galaxy.  While Mg$^0$ atoms are
also likely to reside in the halos of galaxies \citep[e.g.][]{charlton98}, the
expected column density is small and should show significantly
different kinematics.

\subsection{Distance of Neutral Gas to the GRB}
\label{sec:MgI}

Even if physical conditions in the ISM of the GRB host galaxy support
a significant Mg$^0$ column density, the enhanced far-UV radiation
from the GRB afterglow can burn away the gas even at large
distance from the burst.  Observations of Mg$^0$, therefore, can place a
lower limit to the distance of the neutral gas to the burst.

As an estimate to this lower limit,
we calculate the distance $r_{min}$ at which $99.99\%$ of the Mg$^0$
atoms would be ionized by the GRB afterglow.   At this level of
photoionization, one would not expect to detect even the strong
\ion{Mg}{1}~2852 transition.
For an optically
thin slab, $r_{min}$ corresponds to the distance at which the
gas experiences an integrated photon surface
density $N_\gamma = \ln(10^4)/\sigma_{ph}$.  
For a total number of ionizing photons $\phi$, the distance is
\begin{equation}
r_{min}= 100 \times \ltk  \frac{\phi\,\sigma_{ph}}{1.1 \sci{43}} 
\rtk^\ohf \; {\rm pc} \perd
\label{eqn:rlim}
\end{equation}
We adopt the functional form given by \cite{verner96} for the
photoionization cross-section of Mg$^0$, $\sigma_{ph}^{\rm Mg^0}(E)$.
This function peaks at the ionization potential \\
$\sigma_{ph}^{\rm Mg^0}(E=7.64 {\rm eV}) = 1.7 \times 10^{-18} \cm{2}$ 
and declines
sharply with increasing energy.  We assume a value
$\sigma_{ph}^{\rm Mg^0} = 10^{-18} \cm{2}$ (corresponding to $E
\approx 8.5$eV) as a conservative estimate of the cross-section.  For
GRB~051111, we calculate a time-averaged 
luminosity density at $h\nu = 8.5$eV, \\
$<L_\nu>_t = 6.4 \sci{31} \; {\rm erg \, s^{-1} \, cm^{-2} \,
Hz^{-1}}$ over $t=20-1000$\,s, leading to a mean total photon number
$\phi = 2 \sci{60}$ at 8.5 eV.  Substituting $\sigma_{ph}^{\rm Mg^0}$
for $\sigma_{ph}$ in Equation (2) yields $r_{min}= 46$\,pc for the
neutral gas observed in the GRB~051111\footnote{Note that one can ignore
recombination provided that the density is 
not extraordinary, i.e.\ $n_e < 10^9 \cm{-3}$.}.  The presence of 
strong \ion{Mg}{1}
absorption transitions at $t=1500$s after the burst therefore
suggests that the neutral gas is located {\it at least} 50 pc away
from the afterglow.

\begin{table}[ht]\footnotesize
\begin{center}
\caption{Constraints on Circumburst Distances of Observed Neutral Gas\label{tab:rlim}}
\begin{tabular}{lcccccccc}
\tableline
\tableline
GRB & $z$ & $\alpha$ & $\beta$ & Ref & $\log L_\nu^a$
& $r_{MgI}^b$ & $r_{excite}^c$ \\ 
& & & & & (cgs) & (pc) & (pc) \\
\tableline
010222 & 1.477 & 0.80& 0.89&1&31.39 &   40 &  190 \\ 
020813 & 1.254 & 0.85& 0.92&2&31.09 &   30 &  140 \\ 
021004 & 2.328 & 1.05& 1.05&3&32.21 &  140 &  620 \\ 
030323 & 3.372 & 1.56& 0.89&4&32.85 &  540 & 2330 \\ 
030329 & 0.169 & 1.10& 1.00&5&31.38 &   60 &  250 \\ 
050408 & 1.236 & 0.79& 1.30&6&29.93 &   10 &   40 \\ 
050730 & 3.969 & 0.30& 1.80&7&32.16 &   70 &  340 \\ 
050820 & 2.615 & 0.95& 1.00&8&31.97 &  100 &  430 \\ 
051111 & 1.549 & 0.87& 0.60&9&31.32 &   40 &  180 \\ 
060206 & 4.048 & 1.01& 0.51&10&32.41 &  170 &  730 \\ 
\tableline
\end{tabular}
\end{center}
\tablenotetext{a}{Specific luminosity of the GRB afterglow at $t=1000$s for $\nu = 8{\rm eV}/h$.}
\tablenotetext{b}{Radius from the GRB afterglow where 99.99\% of Mg$^0$ 
atoms \\ 
would be ionized.}
\tablenotetext{c}{Approximate radius where a column of $10^{14} \cm{-2}$ Si$^+$ ions would \\
experience one excitation per minute per ion.}
\tablecomments{We caution that the values listed in this Table should \\
be considered rough estimates of the light curves.  \\
The flux of the GRB observed at Earth is given by \\
$F_\nu = (L_\nu / 4 \pi d_L^2 / [1+z]) (t/1000)^\alpha 
(\nu/1.7 \sci{15} {\rm Hz})^{\beta}$.  \\
We have not corrected for the fluxes for dust extinction \\
in the host galaxy.}
\tablerefs{
1: \cite{mhk+02};
2: \cite{bsc+03}; \\
3: \cite{Mir03};
4: \cite{vel+04}; \\
5: \cite{log+04};
6: \cite{foley06}; \\
7: \cite{cpb+05};
8: \cite{gcn3834}; \\
9: \cite{butler06};
10: \cite{danp06}}
\end{table}

A stronger constraint can be placed on the minimum distance from the
GRB by evaluating
time variations in the Mg$^0$ column density \citep{mhk+02}.  In this
case, the result is independent of the total Mg$^0$ column density
prior to the GRB.  For GRB~051111, we set an upper limit to the
decline in $\N{Mg^0}$ over the central 40\kms\ interval to be less
than $25\%$ during a $\Delta t_{\rm obs} = 1800$ s exposure.  The number
of photons emitted over the time interval $t_{obs}=(3840+900)$\,s to
$t_{\rm obs}=(3840+2700)$\,s is $\phi = 2.3\sci{57}$.  Therefore, we
find a lower limit to the separation based on the absence of line-strength
variability to be

\begin{equation}
r_{\rm min}^v = \ltk \frac{\phi \; \sigma_{ph}^{\rm Mg^0}}{4 \pi  
\ln(N_i/N_f)} \rtk^\ohf > \; 80 \, {\rm pc}
\label{eqn:var}
\end{equation}

\noindent for $N_f > 0.75 N_i$.  This is in good agreement
with our estimate for $r_{\rm min}$ from above.  Note that if we were
able to place an upper limit to the variability of $5\%$ then 
$r_{\rm min}^v$ would double.  

We present estimates of $r_{\rm min}$ for several other GRB assuming
Equation~\ref{eqn:rlim}, together with the estimated parameterization
of the light curves shown in Table~\ref{tab:rlim}.  The results for
GRB~020813 and GRB~050408 are notable because those sightlines also
exhibit significant Mg$^0$ gas.  Our analysis demonstrates that the
conditions for GRB~051111 are not atypical.  A generic result of our
analysis is that Mg$^0$ absorption identified in GRB spectra must
correspond to gas at distances $\gtrsim 100$\,pc from the afterglow.

%One serious caveat to the above conclusions 
%is where the ionizing flux is significantly
%attenuated by dust at either $r < 100$\,pc or within the gas cloud
%bearing the Mg$^0$ atoms. [Brief discussion and estimate]
%[Although we use the observed radiation field, one doesn't tend to
%observe to 7eV.  050730 is the obvious exception.]

\subsection{Electron Density}
\label{sec:FeI}

In the previous section, we have argued that the very presence of
Mg$^0$ gas or lack of time variability in its column density
indicate the gas is located at large distance from the GRB afterglow.
In this case, one can examine the gas under the assumptions of
steady-state balance and photoionization equilibrium.  Considering
together the presence (or absence) of atomic species like Mg$^0$ and
Fe$^0$ in a neutral hydrogen gas leads to constraints on the ratio of
the electron density to the intensity of ambient far-UV radiation.
For the ionization balance of Mg$^0$ and Mg$^+$, we express
photoionization equilibrium in terms of the column density ratios
(Appendix~\ref{app:atomic}),
\begin{equation}
n_e = \frac{\N{Mg^0}}{\N{Mg^+}} 
\frac{\Gamma({\rm Mg^0})}{\alpha^{\rm Mg^+}(T)},
\label{eqn:nemg}
\end{equation}
where $\Gamma({\rm Mg}^0)=Const. \times \sigma_{ph}({\rm Mg^0}) \times
G/G_0$ represents the photoionization rate of Mg$^0$ and
$\alpha^{\rm Mg^+}(T)$ is the total recombination coefficient dependent upon
the gas temperature $T$.  Equation (4) demonstrates that
observations of the relative ionization fraction for a single heavy element
constrain $n_e\times (G/G_0)^{-1}$ at a given temperature.

\begin{figure}[ht]
\begin{center}
\includegraphics[height=3.6in,angle=90]{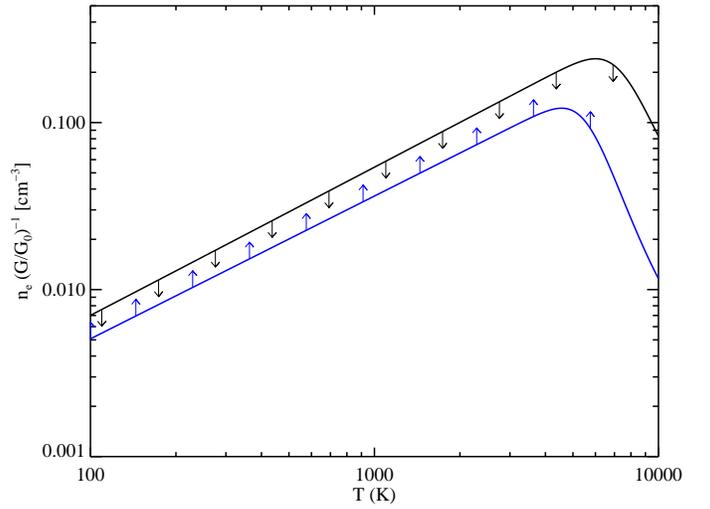}
\caption{Constraints on the ratio of electron density $n_e$
to the ambient far-UV intensity $(G/G_0)$ as a function of the
gas temperature.  The black curve set an upper limit based on
the non-detection of atomic Fe.  The blue curve set a lower
limit based on the detection of atomic Mg and a conservative
upper limit to the column density of Mg$^+$.
}
\label{fig:neTatomic}
\end{center}
\end{figure}

For GRB~051111, the \ion{Mg}{2} doublet is highly saturated in the
echelle data and only a lower limit can be
placed.  We can estimate, however, a conservative 
upper limit to $\N{Mg^+}$ from the
observed column density of Zn$^+$.  First, we assume relative
solar abundances for Zn and Mg.
Next, we increment the observed $\log N({\rm Zn^+})$ by 1\,dex to
account for both line-saturation and a nucleosynthetic 
enhancement relative to solar. 
This treatment presumes the Mg$^0$ gas is
co-spatial with the Mg$^+$ ions inferred from Zn$^+$, which is
supported by the kinematic characteristics of the various line
profiles (Figure~\ref{fig:051111}).  Finally, we derive a conservative
lower limit
to the ratio $\N{Mg^0}/\N{Mg^+} > 10^{-2.8}$.  The corresponding lower
limit of $n_e\times (G/G_0)^{-1}$ as a function of $T$ is shown in the
solid, blue curve in Figure~\ref{fig:neTatomic}.

We repeat this analysis for the observed ratio Fe$^0$/Fe$^+$.  
At $T < 8000$K, radiative recombination dominates and we have
\begin{equation}
n_e = \frac{\N{Fe^0}}{\N{Fe^+}} 
\frac{\Gamma({\rm Fe^0})}{\alpha_r^{\rm Fe^+}(T)} \perd
\label{eqn:nefe} 
\end{equation}
The absence of Fe$^0$ places an upper limit on the column density
ratio $\N{Fe^0}/\N{Fe^+} < 10^{-3}$, which in turn leads to an
upper limit of $n_e\times (G/G_0)^{-1}$ as shown in the solid, black
curve in Figure~\ref{fig:neTatomic}.

%We have shown in Figure 3 that the mean intensity at 1\,kpc due to the
%afterglow is $G/G_0 \approx 10^5$.  Dust extinction would likely
%reduce the radiation field to $G = 10^3-10^4 G_0$.  Assuming $G = 10^3
%%G_0$ and $T< 10000$K, we estimate $n_e = 30 - 400 \cm{-3}$.
%Naturally, the estimated $n_e$ increases if the gaseous cloud is at a
%smaller distance to the afterglow.  Assuming an ionization fraction of
%$x \equiv n({\rm H^+})/n({\rm H}) = 10^{-2}$ implies a hydrogen
%density $n({\rm H}) = 3000 - 4\times 10^4 \cm{-3}$; this value is
%comparable to what is observed in H\,II regions or giant molecular
%clouds in the Galactic ISM.

%[Adopt Le Floch extinction!]
Figure~\ref{fig:neTatomic} allows us to roughly estimate the electron
density of the gas for a known ambient radiation field $G$.  If
the GRB progenitor was located within a star forming region, then the
ambient far-UV radiation field could be large.  Observations of known
GRB host galaxies at $z>1$ show a mean star formation rate integrated
throughout the host ISM of SFR$\sim 10\,M_\odot \, yr^{-1}$
\citep{chg04,lefloch06}.
Even if all
the star formation were concentrated in a single cloud, the mean
intensity at 1\,kpc %from a region of on-going ${\rm
%SFR}=10\,M_\odot \, yr^{-1}$ 
is $G/G_0 \approx 100$.  At $T< 10000$K,
we therefore estimate $n_e = 1-10 \cm{-3}$.  Assuming an ionization
fraction of $x \equiv n_{\rm H^+}/n_{\rm H} = 10^{-2}$, we estimate
the hydrogen volume density $n_{\rm H} = 100 - 1000 \cm{-3}$; this
value is comparable to what is observed in the Galactic H\,II regions
or giant molecular clouds.  
To estimate the size of the neutral gaseous cloud, we adopt the
observed $\log\N{Zn^+} > 13.7$ and assume a solar
metallicity to infer $\log\,\mnhi>21$.  We derive a cloud size of
$l\equiv N_{\rm H}\,/\,n_{\rm H}>3$ pc.  Lower metallicity
would imply a higher \nhi\ and a larger size for the
cloud.  

%Ignoring dust
%extinction, the mean intensity at 1\,kpc from a star forming region
%with SFR=10$\rm M_\odot \, yr^{-1}$ is $G/G_0 \approx 100$.
%Extinction is likely, however, and a more realistic value is $G = 1-10
%G_0$.  Assuming $G = 10 G_0$ and $T< 10000$K, we set an upper limit
%$n_e < 10 \cm{-3}$ and suspect a typical value of $n_e \approx 1
%\cm{-3}$.  Assuming an ionization fraction of $x \equiv n_{\rm
%H^+}/n_{\rm H} = 10^{-2}$, we estimate the hydrogen volume density
%$n_{\rm H} = 10 - 100 \cm{-3}$; this value is typical of cold clouds
%in the Galactic ISM.  [GIVE SIZE]

\section{Implications from Observations of Fine-Structure Transitions}
\label{sec:fine}

Additional constraints on the gas density, UV radiation intensity, and
the spectral shape of the afterglow emission can be learned from
comparing the line strengths of different fine structure lines
\citep[e.g.][]{bw68}.
Although \ion{Fe}{2} and \ion{Si}{2}
fine-structure transitions have yet to be observed in intervening
quasar absorption line systems \citep[e.g.][]{howk05}, they appear to
be a generic feature in the GRB host environment.  Their presence
indicates unique physical conditions in the ISM hosting the GRBs.

The excited states can be produced through either photon pumping or
collisional excitations.  In this section, we will investigate these
various excitation mechanisms.  Specifically, we will consider for
photon pumping (1) direct excitation by IR radiation from sources
local to the gas\footnote{Note that the CMB temperature and photon
density at $z=1.54948$ cannot account for significant external IR
pumping.} and (2) fluorescence following excitations by ultraviolet
photons.  For collisional excitations, we will discuss (3) collisions
with charged particles, primarily electrons, and (4) collisions with
neutral hydrogen.  Although the photon pumping processes are generally
weak in typical astrophysical environments, the radiation field of the
GRB afterglow is extreme (Figure~\ref{fig:radiation}) and can dominate
the excitation processes.

To estimate the strength of the afterglow UV radiation field at the
location of the absorbing gas in GRB~051111, for example,
we adopt the minimum distance
from \S~\ref{sec:MgI} and Equation~\ref{eqn:var},  
$r_{\rm min} = 80$\,pc.  We conclude
that the majority of Fe$^+$ resides at a distance of 80\,pc or greater
from the afterglow.  At $r=80$ pc, the FUV radiation field is
still $>10^7\,G/G_0$ (Figure~\ref{fig:radiation}), 
stronger than that observed for intense star-forming regions.

\begin{figure}[ht]
\begin{center}
\includegraphics[height=3.6in,angle=90]{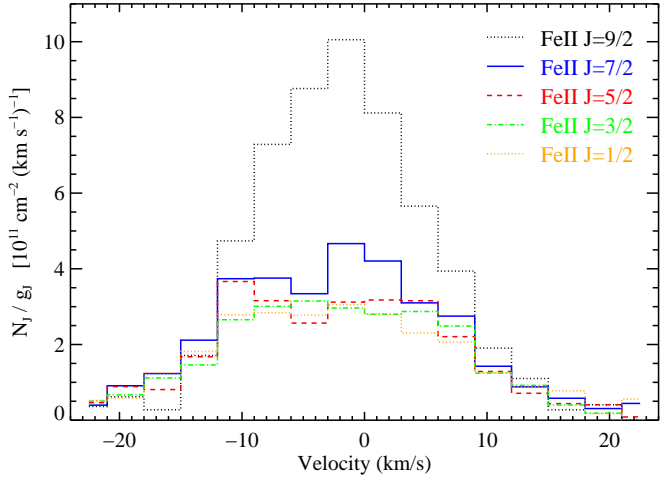}
\caption{Average apparent optical depth profiles normalized
by the degeneracy ($g_J \equiv 2J+1$) of 
the five $^6D_{9/2}$ levels of Fe$^+$ for GRB~051111.  
One notes that the excited states track one another closely but
that the ground state exhibits greater optical
depth at $v > 0 \mkms$ relative to the excited states.
Also note that the $g_J$-normalized column densities of the
$J=9/2$ ground-state level are scaled down by a factor of 20 to
enable comparison with the excited states.
}
\label{fig:hca}
\end{center}
\end{figure}

Figure~\ref{fig:hca} shows the apparent optical depth profiles of the
five $a^6D$ levels of Fe$^+$ from GRB~051111 normalized by their
degeneracy $g_J \equiv 2J + 1$.  We emphasize three observations from
the Figure.  First, the velocity profiles of the excited states (color
histograms) track each other very closely.  Second, the $g_J$
normalized profiles of the excited states have roughly the same value.
Indeed, the total column densities of the $J=7/2$ to $J=1/2$ states
are roughly proportional to the degeneracy $N_J \propto g_J$ with a
modest but statistically significant decrease with increasing $J$.  To
populate the excited levels according to their degeneracy implies a
saturated excitation process, i.e., an excitation rate that is much
faster than the spontaneous decay rate.  Therefore, it is unlikely the
physical conditions vary considerably with velocity and we can
integrate the profiles for the excited states over the entire velocity
range to consider the total column densities. Lastly, the ground state
($J=9/2$) exhibits a discrepant $g_J$ normalized profile from the
excited states, containing $\approx 20\times$ higher column density
than the value expected from the population of the excited levels.
The ground level is clearly not populated according to its degeneracy
relative to the excited states.

Because no viable excitation mechanism will populate only the excited
states according to $g_J$, we conclude that the sightline has
intersected at least two phases of gas: 
one with minimal excitation which
contains roughly $95\%$ of the ground-state gas and one where all of
the $a^6D$ levels are roughly populated according to $g_J$.  These two
phases could be embedded within one another (e.g.\ a dense core within
a less dense cloud) or even represent separate layers of the same
clump.  Figure~\ref{fig:hca} offers additional kinematic evidence for
two distinct phases; one observes that the ground-state profile is
more sharply peaked at $v=0 \mkms$ than the excited states.  For these
reasons, in the following discussion we will not include observational
constraints drawn from the ground states of Si$^+$, Fe$^+$, or O$^0$.
While Figure~\ref{fig:hca} presents only the results for GRB~051111, we note
that the data from GRB~050730 is qualitatively similar; the excited
states are populated according to $g_J$ and the ground state shows a
significant overabundance (Table~\ref{tab:050730}).  
%We suspect this is a generic feature of GRB spectra.

\subsection{Direct IR Pumping}
\label{sec:irpump}

We first examine the population ratios of the excited states under the
scenario of excitation by IR photons.  
%We will find that
%the GRB afterglow could excite the gas provided it is located within
%$\sim 1$\,pc, but we will demonstrate in the next section that the UV
%excitation rate exceeds IR pumping at all distances.  
Direct IR pumping can populate the excited states of Si$^+$, Fe$^+$,
and O$^0$ through forbidden magnetic dipole transitions.  Because
these transitions have selection rules $\Delta J = 0, \pm 1$, excited
states beyond the first level must be populated by an upward cascade
of transitions.  If IR pumping dominates, then the excitation of level
$j$ relative to the ground-state $i$ is given by
\begin{equation}
\frac{N_j}{N_i} = \frac{g_j}{g_i} \frac{N_\nu^\gamma}{1+N_\nu^\gamma}
\end{equation} 
where $N_\nu^\gamma$ is the number of photons at the appropriate
frequency $\nu$
\begin{equation}
N_\nu^\gamma = \frac{I_\nu \lambda^3}{8 \pi h c} \perd
\end{equation}
%
%\noindent We will find that it is possible for the 
%excited states to have been pumped by IR photons
%provided that the GRB afterglow
%is the source of radiation and the gas is located at $r \lesssim 1$pc
%from the burst.  In the following sub-section, however, we will
%demonstrate that UV pumping would dominate at this distance from the 
%afterglow.  We believe this is a generic result.  
%It is very unlikely that IR pumping
%would ever be a dominant excitation mechanism for gas associated with
%a GRB afterglow spectrum because the excitation rates for indirect
%UV pumping will always dominate.

\begin{figure}[ht]
\begin{center}
\includegraphics[height=3.6in,angle=90]{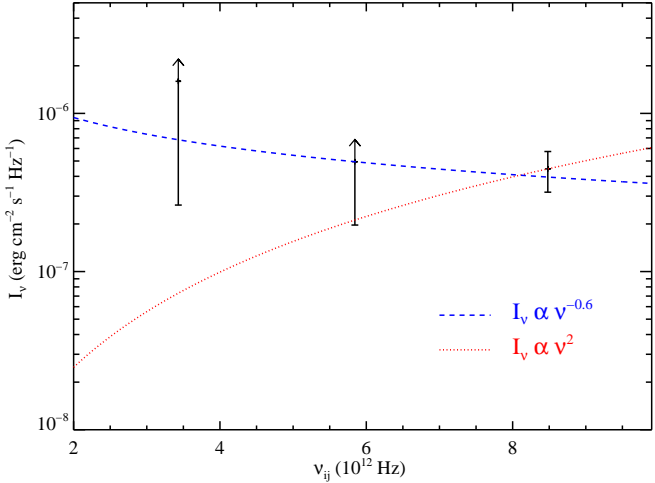}
\caption{Calculated intensity of the radiation field required
to reproduce the relative populations of the $J=7/2,5/2,3/2$ and
$J=1/2$ levels of Fe$^+$ for GRB~051111 via direct IR pumping (magnetic dipole
transitions).  
The error bars reflect 1-$\sigma$ uncertainties in the ratios.  
For the upper two transitions (lower frequency), 
the observed ratios are within $1\sigma$
of satisfying $n_i g_j = n_j g_i$.  Therefore, there is no formal
upper bound to $I_\nu$ and we show the central value but also express
the result as a lower limit.
The blue dashed curve 
qualitatively represents the flux from the GRB afterglow, i.e.\ 
a radiation field with $I_\nu \propto \nu^{-0.6}$.
The red dotted curve shows the
radiation field of a thermal source with $I_\nu \propto \nu^3$.
The latter radiation field is ruled out by the observations.
}
\label{fig:irpump}
\end{center}
\end{figure}

To explore the process of IR pumping, we consider the spectrum of the
radiation field required to reproduce the observed column density
ratios between the excited states of Fe$^+$ for GRB~051111.  We ignore
the $J=9/2$ state because $\approx 95\%$ of the ground-state ions must
arise in a distinct region from the excited Fe$^+$ gas.
Figure~\ref{fig:irpump} presents the intensities inferred from the
observed population ratios of two adjacent levels ($J=7/2\rightarrow
5/2$, $J=5/2\rightarrow 3/2$, and and $J=3/2\rightarrow 1/2$) versus
the corresponding excitation frequency.  The error bars reflect
1-$\sigma$ uncertainties in the ratios.  For the upper two
transitions, the observed ratios are within $1\sigma$ of satisfying
$n_i g_j = n_j g_i$.  Therefore, there is no formal upper bound to
$I_\nu$ and we show the central value but also express the result as a
lower limit.

Superimposed on the plot are the predictions for two radiation fields
with $I_\nu \propto \nu^2$ (e.g.\ a thermal spectrum with $T > 200$K)
and $I_\nu \propto \nu^{-0.6}$ (the optical afterglow spectrum; Butler
et al.\ 2006).  The thermal radiation field is ruled out at high
confidence, whereas the GRB spectrum (or a slightly steeper spectrum, 
$\beta \approx 0$, which might be appropriate for these low frequencies) is
acceptable.  For GRB~050730, where the population of the excited
levels are all consistent with $g_J$ due to large measurement
uncertainties, the data are consistent with a constant $I_\nu >
10^{-6} \; {\rm erg \, s^{-1} \, \cm{-2} \, Hz^{-1}}$ over $\nu =
2-9\times 10^{12}$ Hz.  This is higher than any reasonable IR source
(i.e.\ dust) in even the most intense star-forming regions, and could
only be produced by the GRB afterglow.  

Next, we estimate the number of Fe$^+$ ions that were excited by
IR photons from
the afterglow prior to the start of the echelle observations.  We
calculate the excitation rate as a function of distance from the
afterglow, taking into account the spontaneous decay rate of the
excited levels.  The spontaneous decay coefficient of the first
excited level is $A_{7/2,9/2} = 2.12 \sci{-3}\, {\rm s^{-1}}$ giving a
decay time of 8\,minutes
and an oscillator strength of
\begin{equation}
f_{ji} = \frac{A_{ji} t_0}{\ohf \alpha^3 \frac{g_i}{g_j} E_{ji}^2}
= 2.16 \sci{-8}
\end{equation}
where $t_0 = 2.419 \sci{-17} \rm s$, $\alpha$ is the fine-structure
constant, and $E_{ji}$ is the transition energy (Ryd). 
For a gas `cloud' with column density $\N{Fe^+} = 10^{15} \cm{-2}$ and
Doppler parameter $b=15\mkms$, the optical depth at line center is
\begin{equation}
\tau_0 = \frac{1.5\sci{-2} N \lambda f_{ji}}{b} = 5.6 \sci{-4}
\end{equation}
with all quantities expressed in cgs units.  The estimated optical depth
indicates that the gas is optically thin to the IR photons.  To
estimate the number of photons absorbed by the cloud per decay time,
we calculate the equivalent width in the optically thin limit
\begin{equation}
W_\lambda = 8.85\,\left(\frac{N}{10^{15}\,{\rm cm}^{-2}}\right) \left(\frac{\lambda}{1000\,{\rm \AA}}\right)^2 f_{ji} = 13\,{\rm m\AA} \perd
\end{equation}

Figure~\ref{fig:radiation} shows the number of IR photons emitted from
$t_{\rm obs}=50$ s to 3840 s with energy $E_{9/2,7/2} = 385 \cm{-1}$
in a 1\AA\ window.  The number of excitations per ion\footnote{Note
that the rate is independent of the number of ions assumed here
provided the optical depth remains small.} per decay time is
\begin{equation}
n_{excite}^{IR} = \left(\frac{W_\lambda}{1\,\rm\AA}\right) \left(\frac{N_\nu^\gamma}{N_{\rm ion}}\right) \left(\frac{\Delta\,t_{\rm decay}}{\Delta\,t}\right) \ltp \frac{r}{1\,\rm pc} \rtp^{-2} = 26 \ltp \frac{r}{\rm pc} \rtp^{-2},
\label{eqn:irrate}
\end{equation}
where $N_\nu^\gamma = 10^{18.8}$ photons per cm$^{-2}$, $n_{\rm
ion}=10^{15}$ ions per cm$^{-2}$, $\Delta\,t_{\rm decay}=480$ s,
$\Delta\,t=\Delta\,t_{\rm obs} / (1+z)=1487$ s.

According to Equation~\ref{eqn:irrate}, there is a sufficient photon
flux at $r=1$\,pc from GRB~051111 to excite the 
Fe\,II $J=7/2$ level prior to the onset 
of the echelle observations but the photon flux is too low at $r=80$\,pc.
Our analysis therefore indicates that IR
pumping is unlikely the primary mechanism for producing the excited
ions found in GRB~051111.
Furthermore, in the next sub-section we will show
that the excitation rate by UV pumping is orders of magnitude higher
than IR pumping.  Therefore, we conclude 
that {\it IR pumping is an unimportant excitation mechanism for
gas observed in GRB afterglow spectra}.

%Regarding the GRB intensity, an extrapolation of
%Equation~\ref{eqn:flux} to IR frequencies indicates $I_\nu^{GRB} >
%10^{-7} {\rm erg \, s^{-1} \, \cm{-2} \, Hz^{-1}}$ for $r < 20$ pc at
%50\,s/$(1+z_{GRB})$ in the GRB host galaxy (see
%Appendix~\ref{app:cosm}).  Although the IR intensity of the GRB is
%considerable, the more relevant issue is whether the afterglow would
%pump a non-negligible number of Fe$^+$ ions prior to the start of the
%observations.  

%  We note, however, that
%Fe$^+$ could not have survived the intense FUV radiation field at this
%close distance.  the intensity at 1 pc is far in excess of the
%intensity inferred from the observations (Figure~\ref{fig:irpump}).
%At $r=10$ pc, the intensity is roughly consistent but the excitation
%rate is less than one per ion per decay time.  It is unclear,
%therefore, whether one can sufficiently excite the Fe$^+$ ions without
%driving them to a population strictly according to $g_J$ which for
%GRB~051111 violates the observed ratio of the $J=7/2$ to $J=5/2$
%levels.

%A full analysis of IR pumping by the GRB afterglow
%requires a treatment of the time dependence of excitation
%which is beyond the scope of this paper.
%For now we note that IR pumping is a potentially viable process.
%We emphasize, however, that we will find in the next
%sub-section
%that the excitation rate by UV pumping is orders of magnitude higher
%than IR pumping.  Therefore, we conclude 
%that IR pumping is an unimportant excitation mechanism for
%gas in GRB afterglow spectra.

\subsection{Indirect UV Pumping}
\label{sec:uvpump}

The fine-structure states of Fe$^+$, Si$^+$, and O$^0$ ions can also
be populated by the absorption of a UV photon to an upper level,
followed by spontaneous radiative decay to an excited level of the
ground term.  Requiring electric dipole transitions, the maximum
angular momentum change is $|\Delta J_{max}| = 2$.  Therefore, one
requires a series of excitations to populate the $J=3/2, 1/2$ levels
of Fe$^+$, and only a single absorption for the Si$^+$ ion and O$^0$
atom.  The analysis of UV pumping is similar to IR pumping, although
it requires a careful consideration of all the upper levels.  To
perform the calculation, we adopt the software packaged Popratio
developed by \cite{silva01,silva02} which incorporates modern atomic
data for these various ions.

Similar to the treatment of IR pumping, we investigate (1) whether the
observations describe a physically sensible radiation field; and (2)
whether the pumping rate is sufficient to excite a significant column
of gas prior to the start of spectroscopic observations assuming the GRB
afterglow is the radiation source.  We first examine the radiation
field required to explain the observed population ratio of different
excited states.

\begin{figure}[ht]
\begin{center}
\includegraphics[height=3.6in,angle=90]{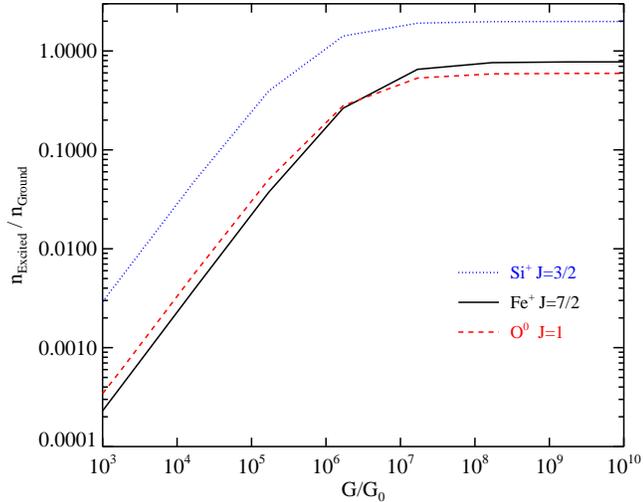}
\caption{Indirect UV pumping of the first excited state
relative to the ground state for 
Si$^+$, Fe$^+$, and O$^0$ versus the far-UV intensity (normalized
to Habing's constant, $G_0$). 
At large intensity, the levels saturate to ratios according to their
relative degeneracies $g_J$.  Note that Si$^+$ exhibits a much higher
excitation ratio than the other gas.
}
\label{fig:uvpump}
\end{center}
\end{figure}

Figure~\ref{fig:uvpump} shows the excitation of the first excited
states of Fe$^+$, Si$^+$, and O$^0$ relative to the ground state as a
function of the far-UV intensity in terms of Habing's constant $G_0 =
1.6 \sci{-3} \, {\rm erg \, cm^{-2} \, s^{-1}}$.  The results are
drawn from Popratio calculations assuming physical characteristics for
the gas such that UV pumping dominates (i.e.\ $n_e = 1 \cm{-3}$).  We
emphasize three points from the figure.  First, all of the excited
states saturate at $G/G_0 > 10^7$.  At these intensities, the levels
are populated only according to the degeneracy of the excited state
relative to the ground state and the differences between ions reflect
the various $J$ values.  Second, the excitation fraction is negligible
and is nearly proportional to $G$ for $G/G_0 < 10^5$, beyond which
pumping induced excited ions become dominant.  Finally, Si$^+$
exhibits $\approx 10$ times higher excitation fraction than Fe$^+$ and
O$^0$ for the same $G/G_0$ values.  This is partly due to the fact
that the \ion{Si}{2} levels are inverted ($g_J$ for the excited state is
twice that of the ground state), but also arises from the detailed
atomic physics of the upper levels\footnote{It is not an effect of UV
absorption strengths; all of the ions exhibit a significant number of
resonance transitions in the far-UV.}.

If we adopt a minimal excitation for Fe$^+$, then for \\
GRB~051111 only
roughly 7\% of the observed $\nfej{9/2}$ originates in the excited
phase, then our observations imply a UV radiation field of $G/G_0>
10^{7}$ under a steady-state assumption (Figure~\ref{fig:uvpump}).  At this
radiation intensity, we also infer a population ratio for Si$^+$,
$n_{J=3/2}/n_{J=1/2}\sim 2$, indicating that roughly 5\% of the
observed ground-state Si$^+$ originates in the excited phase.  This
exercise shows that roughly the same fraction of the observed Fe$^+$
and Si$^+$ ions have been excited toward GRB~051111 under the 
indirect UV pumping mechanism.

%We
%note that the difference between Si$^+$ and Fe$^+$ can be understood
%for the following reasons.  First, the number of transitions available
%to excite Fe$^+$ is at least a factor of two more than those for
%Si$^+$.  Second, at $\N{Si^+}) \ge 10^{15.8} \cm{-2}$ the majority of
%Si\,II transitions are optically thick to UV radiation, making UV
%pumping much less efficient.  Finally, the spectral energy
%distribution of GRB afterglow at UV wavelengths (Equation 1) implies a
%higher photon flux at $\lambda > 2200$ \AA, where many of the Fe\,II
%transitions lie, than at shorter wavelength, where all the Si\,II
%transitions arise.  These factors together reduce the total UV pumping
%rate for the Si$^+$ ions.
%resulting in a smaller fraction of the ground state Si$^+$ to originate
%in the excited phase.

Examining the far-UV radiation field of a GRB afterglow
(Figure~\ref{fig:radiation}), we note that the unattenuated intensity
is sufficient in steady-state conditions to excite all of the ions
out at even several hundred parsecs where $G/G_0 > 10^6$.  In
particular, the luminosity of the GRB exceeds that for star-forming
regions with SFR=100\,${\rm M_\odot \, yr^{-1}}$ by a factor of 1000.
Therefore, if the excitation rate of the ions is large on the
time-scale of the afterglow observations (e.g.\ minutes to hours), then
the burst should excite Fe$^+$, Si$^+$, and O$^0$ via indirect UV
pumping.

The excitation rate due to fluorescence transitions can be roughly
estimated under the assumption that the gas is optically opaque to
many of the resonance transitions.  In contrast to IR pumping where
the process occurs in a cascade of magnetic dipole transitions, UV
pumping proceeds through many transitions ($\approx 10-20$) from the
ground-state multiplet to upper-level multiplets.
%  Also in contrast to IR pumping, many
%of the resonance line transitions have very large optical depths for
%the conditions observed in the ISM of GRB host galaxies.  Therefore, a
A detailed radiative transfer calculation is necessary given the
transient nature of the GRB.  Such a calculation is beyond the scope
of this paper and will be the subject of a future work (Mathews \&
Prochaska, in prep).  Here we present a simple framework for such
calculations.

To estimate the number of far-UV photons absorbed per decay time, we
first consider Si$^+$.  Adopting a column density $\N{Si^+} =
10^{14.5} \cm{-2}$ and Doppler parameter $b=10\mkms$ which roughly
matches the profile of the excited gas for GRB~051111, we first
produce a model spectrum of the ground-state
\ion{Si}{2} transitions with $912\,{\rm \AA} <
\lambda_{\rm rest} < 2000\,{\rm \AA}$ assuming the atomic data given in
\cite{morton03}.  We then measure the equivalent width
of each \ion{si}{2} transition which allows us to
calculate the number of photons
absorbed per second from the GRB afterglow spectrum 
(Equation~\ref{eqn:flux}).   
Summing over all of the transitions and evaluating the GRB afterglow
at $t_{obs}=50$s and $r=10$\,pc, 
we calculate the total number of photons absorbed per minute per ion
%(analogous to Equation~\ref{eqn:irrate}):
\begin{equation}
n_{excite}^{UV} = 80 \ltp \frac{<L_{\nu = 7 \rm eV}>}{5.4 \sci{31}} \rtp
\ltp \frac{r}{10\,\rm pc} \rtp^{-2} 
\ltp \frac{N_{ion}}{10^{14.5} \cm{-2}} \rtp^{-1}
\; {\rm min^{-1}} \; {\rm ion^{-1}},
\label{eqn:siuv}
\end{equation}
assuming the unattenuated spectrum given by Equation~\ref{eqn:flux}.
Although only a fraction of these photons will excite the Si$^+$ ions
(roughly $66\%$), the rate is sufficiently high to show a modest
\ion{Si}{2}* column density at $r=50$ pc at $t = 20$ min after the
burst.  Even at $r=100$ pc, there would be $\approx 16$ excitations
per ion prior to the onset of observations for GRB~051111.  Therefore,
we consider the excitation rate to be sufficient out to at least
$r=100$\,pc from this afterglow.  

We carry out similar calculations for O$^0$ and Fe$^+$ and find
results that match Equation~\ref{eqn:siuv} to within a factor of 3
(Fe$^+$ has a higher rate and O$^0$ is lower, primarily reflecting the
rest wavelengths of the resonance transitions).  We emphasize that
these excitation rates exceed IR pumping by more than an order of
magnitude (compare Equations~\ref{eqn:irrate} and \ref{eqn:siuv}).
Therefore, we conclude that indirect UV pumping is the principal
excitation mechanism in GRB when photon pumping dominates.

We note that the excitation rate per ion given by
Equation~\ref{eqn:siuv} is inversely proportional to the number of
ions assumed.  This is because the gas is optically thick in the
majority of \ion{Si}{2} transitions.  The same holds for \ion{Fe}{2}
and \ion{O}{1} transitions.  Therefore, the equivalent width (hence,
the number of photons absorbed) is insensitive to the column density
assumed provided $N_{ion} \gtrsim 10^{14} \cm{-2}$.  Given the
transient nature of the GRB, this has important implications for the
excitation of these ions.  Most importantly, for a gas cloud (or
clouds) with total column density $N_{ion} > 10^{15} \cm{-2}$, only a
fraction of the gas along the sightline is excited prior to the onset
of observations.

To zeroth order, the number of excited ions is given by the
column density where the majority of transitions reach the
saturated portion of the curve-of-growth.  We have investigated
this column density for the Fe$^+$, Si$^+$, and O$^0$ gas.
We find that for $b=10\mkms$, the excitation saturates at
$N_{ion}^{lim} \approx 10^{14.5} \cm{-2}$ for Si$^+$ and Fe$^+$ and 
a few times larger value for O$^0$.  To first order, this 
limiting column density scales with the Doppler parameter of the
gas, i.e.\ wider profiles should show larger column densities
of excited gas when the excitation rate is limited by line saturation.
We suspect that is not a coincidence that the column densities of
excited gas in GRB~051111 and GRB~050730 are roughly consistent
with this calculation.
%We stress that for the former GRB, in particular, we would
%have been sensitive to $10\times$ lower column density than
%observed because many of the excited \ion{Fe}{2} transitions
%are saturated.

In summary, UV pumping is a viable mechanism for excitations of the
fine-structure states observed in GRB afterglow spectroscopy.
For the case of GRB~051111, we find the gas would be pumped 
to at least 100\,pc from the GRB afterglow 
if the far-UV light is not heavily attenuated by dust
or intervening metal-rich gas.  We draw similar conclusions for 
GRB~050730 and the other GRB listed in Table~\ref{tab:rlim}.
In short, the radiation field of GRB afterglows is likely to 
excite Si$^+$ to at least 100\,pc in the majority of systems
observed to date.  

\begin{figure}[ht]
\begin{center}
\includegraphics[height=3.6in,angle=90]{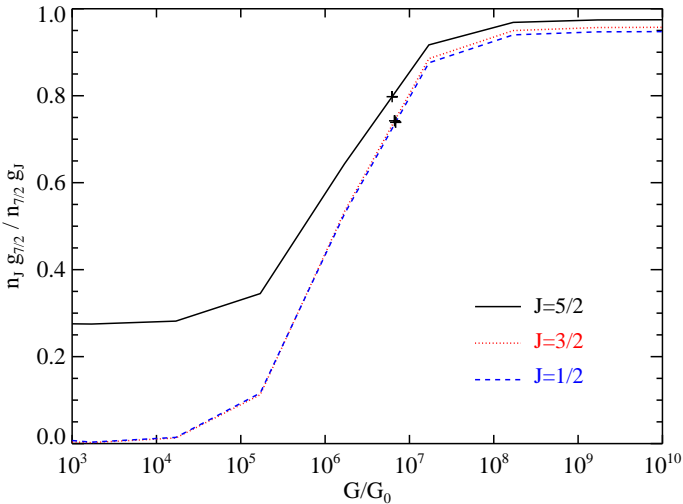}
\caption{Indirect UV pumping of the Fe$^+$ excited levels relative to the
$J=7/2$ level 
(normalized by the degeneracy of the states $g_J$)
as a function of the far-UV intensity. 
The levels saturate at high intensity, 
i.e., they are populated according to $g_J$. 
Also note that the $J=3/2$ and $J=1/2$ levels track one another
at all intensity.  This occurs because the two states require the
absorption of two photons to be populated via UV pumping.  The `+'
signs indicate the observed ratios of the excited levels for the
gas toward GRB~051111.  These observations are consistent with
an intensity $G/G_0 \approx 10^{6.7}$.
}
\label{fig:pumpfe}
\end{center}
\end{figure}

While we have demonstrated that GRB afterglows are sufficiently
bright to UV pump gas to large distance on short time-scales,
we must also consider whether the relative populations of the 
excited states reproduce the observed ratios.  In the next 
section, we will demonstrate that the \ion{Fe}{2} excited levels for
GRB~051111 are well described by a Boltzmann distribution with
excitation temperature
$T_{Ex} = 2600$K.  Because UV pumping is not a thermal process,
one does not expect a Boltzmann distribution aside from the trivial
case where the levels are saturated corresponding to
$T_{Ex} \gg E_{J=1/2}/k$.  
Indeed, this is the case for GRB~050730.
In Figure~\ref{fig:pumpfe}, we present the predicted abundances
of the $J=5/2, 3/2$ and 1/2 levels relative to the $J=7/2$ state
normalized by the degeneracies $g_J$ as a function of the 
far-UV intensity.  
In contrast with IR pumping, 
the spectral shape is unimportant to this calculation;
the results only depend on the overall excitation.

An important prediction for UV pumping is that the $J=3/2$ tracks the
$J=1/2$ level for all intensities.  At $G/G_0 > 10^8$, the levels are
saturated, i.e.\ populated according to $g_J$.  At lower intensity
$G/G_0 < 10^6$, the relative populations are strictly inconsistent
with a Boltzmann distribution.  For intermediate values ($G/G_0
\approx 10^{6.7}$), however, the differences from a Boltzmann
distribution are small.  The plus signs marked in
Figure~\ref{fig:pumpfe} show the observed ratios for GRB~051111 and
indicate the data are well fit by a pumping scenario with a single
intensity.

To conclude, we find that UV pumping can reproduce a Boltzmann
distribution of \ion{Fe}{2} levels with $T_{Ex} > 2000$K.  A key prediction
of UV pumping is that the relative population of the $J=3/2$ and 1/2
levels is given by the ratio of their degeneracies: $n_{3/2}/n_{1/2} =
2$.  This is inconsistent with a Boltzmann distribution for $T_{Ex} <
1000$ K and is a direct test for distinguishing between collisional
excitation and UV pumping: the observation of $n_{3/2}/n_{1/2} > 2$
would rule out UV pumping.  In the cases of GRB~051111 and GRB~050730,
however, we find that $n_{3/2}/n_{1/2} \approx 2$ and that both UV
pumping and collisional excitation are viable mechanisms.

\begin{figure}[ht]
\begin{center}
\includegraphics[height=3.6in,angle=90]{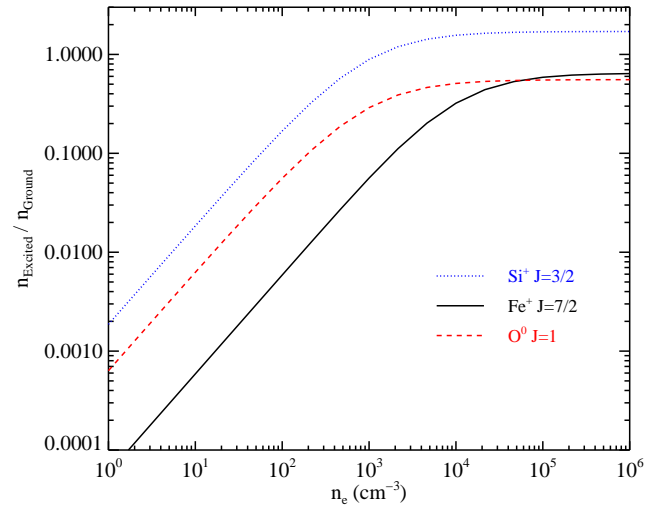}
\caption{Collisional excitation of the first excited states 
relative to the ground state for 
Si$^+$, Fe$^+$, and O$^0$ versus the electron density.  Note
that for O$^0$, 
neutral collisions dominate
and we have assumed an ionization fraction $x=10^{-2}$.
}
\label{fig:collisions}
\end{center}
\end{figure}

\subsection{Collisional Excitation}

The final excitation mechanism to consider is collisional excitation.
For the Fe$^+$ and Si$^+$ ions, collisions with electrons are likely
to dominate the rates\footnote{We caution that the neutral collision
rates for Fe$^+$ have not yet been precisely calculated or measured in
the laboratory.}.  For O$^0$, the rates for collisions with hydrogen
atoms are much faster and are dominant for a mostly neutral cloud,
i.e., $x<10^{-2}$.  The collision rates do vary with temperature but
by less than an order of magnitude for temperatures $T = 100$ to
10000 K.  In the following, we will present results for $T=2600$ K.

In Figure~\ref{fig:collisions} we present the population
the first excited state relative to the ground state as a function of
electron density $n_e$ (neutral hydrogen density $n_{\rm HI}$ for
O$^0$).  The results are qualitatively similar to those for UV pumping
(Figure~\ref{fig:uvpump}) where one exchanges density with far-UV
intensity.  At large density, the excited states are populated
according to their relative degeneracies although reduced by the
Boltzmann function (see Equation~\ref{eqn:boltz} below).  
At low density, the population is
proportional to the density of the colliding particle.  Also, the
excitation of Si$^+$ occurs at significantly lower density although
note that the excitation of O$^0$ would be comparable for gas with low
ionization fraction.  In contrast to UV pumping, the excitation of
Fe$^+$ requires significantly higher densities than Si$^+$ and O$^0$.
One finds $n_e > 10^5 \cm{-3}$ for even a modest population of the
$J=7/2$ level.

When the collisional de-excitation rate $q_{ji} n_e$ exceeds the
spontaneous decay rate $A_{ji}$, the excited states are populated
according to a Boltzmann distribution,
\begin{equation}
\frac{n_i}{n_j} = \frac{g_i}{g_j} \exp \ltk -(E_{ij}/k) / 
 T_{Ex} \rtk \cmma
\label{eqn:boltz}
\end{equation}
where $E_{ij} \equiv E_i - E_j$ is the difference in energy between
the two states, and $T_{Ex}$ is the excitation temperature.  Comparing
predictions following Equation~\ref{eqn:boltz} 
with the observed column densities
of ions at different excited states allows us to examine whether
collisional equilibrium is a representative model for the excitation
mechanism.  A consistent fit in turn leads to a strong constraint for
the gas temperature.

\begin{figure}[ht]
\begin{center}
\includegraphics[height=3.6in,angle=90]{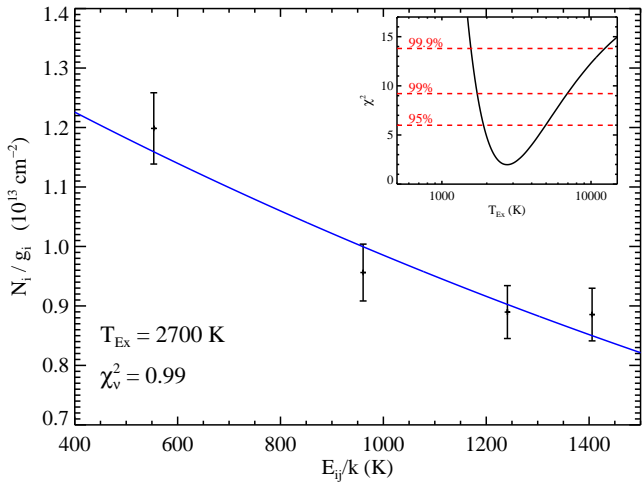}
\caption{Plot of $N_i/g_i$ for the Fe$^+$ excited 
states toward GRB~051111 
as a function of their energy $E_i$ above the ground state.
Overplotted on the figure is the function $A \exp[-(E_{ij}/k)/T_{Ex}$
for a minimum $\chi^2$ value of $T_{Ex} = 2700$\,K.
The inset shows the reduced $\chi^2$ values for the best fit for
a range of $T_{Ex}$ values.
}
\label{fig:boltz}
\end{center}
\end{figure}

Figure~\ref{fig:boltz} shows for GRB\,051111 the observed column
densities scaled by the corresponding degeneracy $N_i/g_i$ for the
\ion{Fe}{2} excited states as a function of the energy $E_{ij}$ above
the ground state $J=9/2$.  The error bars reflect 1-$\sigma$
uncertainty in $N_i$.  The modest decrease of $N_i/g_i$ with $E_{ij}$
suggests the levels are Boltzmann populated with an excitation energy
greater than the largest $E_{ij}$ value.  The solid (blue) curve
indicates the best-fit Boltzmann function with a best-fit excitation
temperature $T_{\rm Ex}=2600$ K.  The inset shows the minimum $\chi^2$
values as a function of $T_{\rm Ex}$; the $99\%$ c.l. limits are
$T_{\rm Ex}(99\%) = 1800 - 6000$ K.  The reduced $\chi^2$ is nearly
unity, $\chi_\nu^2=0.99$, supporting the assumption that the Boltzmann
function is a representative model.

\begin{figure}[ht]
\begin{center}
\includegraphics[height=3.6in,angle=90]{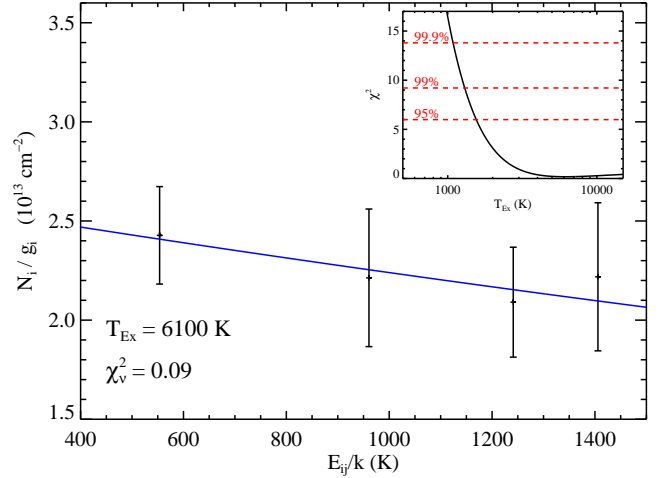}
\caption{Same as Figure~\ref{fig:boltz} except for
GRB~050730.
}
\label{fig:bz050730}
\end{center}
\end{figure}

We have repeated this analysis for GRB\,050730 and the results are
presented in Figure~\ref{fig:bz050730}.  The column density
measurements have larger uncertainty and therefore place weaker
constraints on $T_{Ex}$. Nevertheless, the Fe$^+$ excited state
population is well fit by a Boltzmann function with $T_{\rm Ex}
\approx 6000$K with a firm lower limit of 1000K.

\subsection{Summary}

%1.  Saturation makes it difficult to distinguish mechanisms
%2.  All 3 mechanisms can contribute although IR pumping is minor
%3.  UV pumping will dominate for $r < 10$pc.  Optical depth effects
%need further exploration.
%4.  Small dynamic range for detection without saturation??

After examining the various mechanisms for exciting \ion{Fe}{2}, 
\ion{Si}{2} and \ion{O}{1}
excited levels in the presence of a GRB afterglow, we
have demonstrated that indirect UV pumping and collisional excitation
are both viable mechanisms for producing the excited ions.  For indirect
UV pumping to work, it requires an intensified far-UV radiation field:
$G/G_0 > 10^6$.  For collisional excitation to dominate, it requires
a dense environment of $n_e > 10^3\, {\rm cm}^{-3}$.  We 
rule out IR pumping for gas at distances $\gtrsim 10$\,pc and also
find that the excitation rate is small compared to UV pumping at all
distance (compare Equations~\ref{eqn:siuv} and
\ref{eqn:irrate}).  Therefore, the excitation of gas near GRB
afterglows is primarily determined by indirect UV pumping and/or
collisional excitation.

\begin{table*}[ht]\footnotesize
\begin{center}
\caption{{\sc TESTS OF UV PUMPING AND COLLISIONAL EXCITATION\label{tab:tests}}}
\begin{tabular}{lll}
\tableline
\tableline
Observation & UV Pumping & Collisions \\
\tableline
$\nfej{3/2} = 2 \nfej{1/2}$ & Required & $n_e > 10^4 \cm{-3}$, $T>2500$K \\
Inverted Fe$^+$ population & Ruled out & $n_e < 10^3 \cm{-3}$ \\
Variability & Expected & Ruled out \\
Absence of atomic gas & Expected & Challenging \\
Overabundance of the Fe$^+$ $J=9/2$ level & Expected & Challenging\\
$[\nfej{7/2} - \nsij{3/2}] < [-1 + \epsilon({\rm Fe/Si})]^a$ & 
Ruled out & Allowed \\
$\noij{1} < \nsij{3/2}$ & Ruled out & Allowed for $x > 10^{-1}$ \\
$N_{excited} > 10^{15} \cm{-2}$ & Pre-burst only & Allowed \\
\tableline
\end{tabular}
\end{center}
\tablenotetext{a}{$\epsilon({\rm Fe/Si})$ is the intrinsic, logarithmic Fe/Si gas-phase 
abundance allowing for nucleosynthetic and differential depletion effects.}
\end{table*}

Another important conclusion is that the predictions for UV pumping
and collisional excitation are very similar.  For example, there is
the trivial case at very high intensity or density where the excited
levels are principally populated according to their degeneracy $g_J$,
modified by the Boltzmann factor collisions.  Although the detailed
level populations are different because of the Boltzmann factor, one
requires very high SNR observations to distinguish the cases if
$T_{Ex} > 2000$K.  In the next section, we will consider a few
additional tests for distinguishing between the excitation mechanisms.

Regarding the gas near GRB afterglows, it is clear from
Table~\ref{tab:rlim} that photoionization dominates at distances less
than 10\,pc from the GRB and radiative processes are significant to
100\,pc or further for the majority of afterglows.  We conclude that
circumstellar gas associated with the GRB progenitor will be excited
by UV photons, if the gas is not entirely ionized by the event.
Therefore, it is difficult to determine the density and temperature of
such gas from the analysis of fine-structure transitions.  A full
calculation of the optical depth and time dependent processes of
excitation are warranted and could help reveal the physical conditions
in the absorbing medium (Mathews \& Prochaska, in prep).

\section{Distinguishing Between Photon Pumping and Collisional Excitation}
\label{sec:disting}

In the previous section, we argued that the excitation of
fine-structure states of Fe$^+$, Si$^+$, and O$^0$ gas can occur along
the sightline to GRB afterglows because of indirect UV photon pumping
and/or collisional excitation.  Each process has different
implications for the physical conditions of the gas (e.g.\ density,
temperature).  It is therefore important to distinguish between the
two mechanisms.  We summarize a set of tests in Table~\ref{tab:tests}.

There is a sufficiently large number of \ion{Fe}{2} transitions
covering a wide range of UV wavelengths in the rest frame and nearly
every GRB afterglow spectrum includes the full set of \ion{Fe}{2}
excited levels (c.f.\ GRB~050730 and GRB~051111).  Those
sightlines which exhibit positive detections of the excited states
allow different tests of the excitation mechanism; we consider the two
cases separately.

\subsection{When Excited States of Fe$^+$ are Detected}

%When one detects gas populating the fine-structure levels of Fe$^+$, 
%the relative populations allow for specific tests of the excitation
%mechanism.  Figure~\ref{fig:pumpfe} shows that if UV
%pumping is dominant, then the relative abudnance of the
%$J=3/2$ and $J=1/2$ states will
%be set by their degeneracy, i.e.\ 
%$\nfej{3/2} = 2 \nfej{1/2}$.  Therefore, an absorption system which
%contradicts this relation is very likely to be gas predominantly excited by
%collisions.  It follows that there is only a restricted range of
%density and temperature which would give $\nfej{3/2} \approx 2 \nfej{1/2}$
%from collisions.  Specifically, one would require $n_e > 10^4 \cm{-3}$ and
%$T > 2500$K.

We have shown that under excitation equilibrium the indirect UV
pumping scenario predicts a population ratio that is proportional to
the degeneracy of the excited level, while the collisional excitation
scenario predicts a population ratio that follows the Boltzmann
function.  In principle, one can therefore distinguish between these
two excitation mechanisms based on the observed relative abundances
between different excited ions.  Our analysis has shown, however, that
measurement uncertainties often prohibit us from achieving the goal.
Despite the caveat in obtaining more precise measurements, we highlight 
in this section two features that supports photon pumping and two features
that supports collisional excitation for future studies.

Under photon pumping, we first note in Figure~\ref{fig:pumpfe} 
that the $J=5/2$ state
is significantly more populated than the higher levels at low
intensity $G/G_0 < 10^5$.  This can be understood by the fact that one
can populate the $J=5/2$ and $J=7/2$ states with a single photon
whereas excitation to the $J=3/2$ and $J=1/2$ levels requires at least
two-photon processes.  Second, detection of line variability offers
another discriminator between UV pumping and collisional excitation
processes.  For collisional excitation, we do not expect variability
on time-scales that could conceivably be probed by GRB afterglows
(i.e.\ less than a few days).  In contrast, the absorption-line
strength is expected to vary under photon pumping, because (1) Fe$^+$
becomes more ionized with time due to the afterglow; (2) the afterglow
radiation field declines with time, leading to decreasing $G/G_0$; and
(3) the diffusion of photons through an optically thick cloud is a
time dependent process and could lead to both higher and lower column
densities of the excited levels (Mathews \& Prochaska 2006).  To
observe both features require that additional echelle observations be
carried out at a later time.

We also note that in one extreme there is a maximum column density
$N_{max}^{UV}$ of ions that can be excited by UV pumping prior to the
onset of spectroscopic observations, because of optical depth effects.
The value of $N_{max}^{UV}$ is primarily established by the velocity
structure of the medium (i.e.\ the relative velocity of the absorbing
clouds and their Doppler parameters) and to a lesser extent the size
of the cloud ($1/r^2$ dimming).  An accurate estimate of
$N_{max}^{UV}$ will require a time dependent radiative transfer
calculation, but we note that nearly all of the \ion{Fe}{2},
\ion{O}{1}, and \ion{Si}{2} transitions saturate at $N_{max}^{UV}
\approx 10^{15} \cm{-2}$.  Therefore, if we observe a column density
of excited gas significantly in excess of $10^{15} \cm{-2}$, this
would be strong evidence for collisional excitation, or possibly
pre-burst UV pumping as optical-depth effects would not prevent
the entire cloud from being excited \citep{sarazin79}.

%First, the gas exhibiting Fe$^+$ fine-structure
%could be photoionized by the GRB afterglow.
%This would lead to lower column densities in time, 
%especially during the first few hours of the burst.  
%Second, the fading of the GRB afterglow implies a 
%lower UV pumping rate at late times, e.g., one might
%observe an afterglow at early and late times when the
%luminosity differs by a factor of 100.  Because the Fe$^+$ levels 
%have short decay times ($\approx 8$min for
%the $J=7/2$ to $J=9/2$ transition), one could witness a decline 
%in the population
%of Fe$^+$ excited states.  Finally, it is possible that optical
%depth effects would give line-strength variability.  Specifically, the 
%diffusion of photons through an optically thick cloud is a time dependent
%process and could lead to both higher and lower column densities
%of the excited levels (Mathews \& Prochaska 2006).
%All of these examples of line-strength variability 
%are associated with photon pumping
%and would mark this as the dominant excitation mechanism.

Under collisional excitation, we first note that at the implied gas density
$n_e > 10^3 \cm{-3}$ (Figure~\ref{fig:collisions}) we also expect to
observe large fractions of the elements C, Fe, and Mg in their atomic
states.  Although we have shown in $\S$~\ref{sec:MgI} that the GRB
afterglow can photoionize atomic gas at large distances, there is a
relatively narrow window where the gas would be ionized and
collisional excitation would dominate indirect UV pumping.  Therefore,
the presence of Fe$^0$, Mg$^0$, or C$^0$ may be viewed to support
collisional excitation.

\begin{figure}[ht]
\begin{center}
\includegraphics[height=3.6in,angle=90]{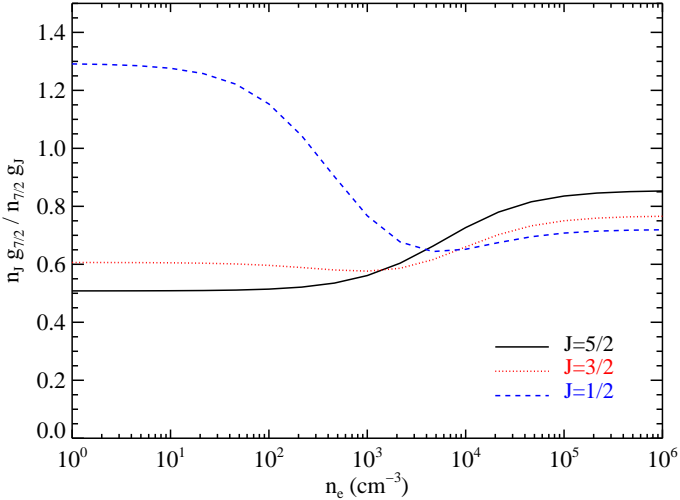}
\caption{Collisional excitation of the \ion{Fe}{2} excited levels 
relative to the $J=7/2$ level 
normalized by the degeneracy of the states $g_J$
as a function of the electron density.
The levels saturate at high density,
i.e., they become populated according to $g_J$ weighted
by the Boltzmann factor $\exp(-\Delta E/ kT_{Ex})$ where we have assumed
$T_{Ex} = 2600$K.
Also note that at intermediate density the relative populations are
roughly the same and that at low density the $J=1/2$ and $J=3/2$
exhibit an inverted population.
}
\label{fig:collfe}
\end{center}
\end{figure}

Second, we note in Figure~\ref{fig:collfe} 
that under collisional excitation for a
gas temperature $T_{Ex} = 2600$ K the population ratios between
different excited levels are significantly different from the
Boltzmann distribution at $n_e<10^3\,{\rm cm}^{-3}$.  Specifically,
both $J=5/2$ and $3/2$ levels are significantly underabundant relative
to the $J=7/2$ level.  In addition, the population becomes inverted,
especially the $J=1/2$ level.  The relative abundance at these lower
densities contrast with anything predicted for UV pumping.  Although
Figure~\ref{fig:collfe} applies to a single phase of gas with $T=2600$
K, the results are similar for any temperature $T < 10000$ K.  It is,
however, very challenging to observe an inverted population of \ion{Fe}{2}
levels because the absolute excitation rate of Fe$^+$ is small for
$n_e < 10^3 \cm{-3}$.

An intriguing aspect of the \ion{Fe}{2} excited levels for
GRB~051111 and GRB~050730 is the overabundance of the ions in the
ground level.  We also observed for GRB~051111 that the ground-state
of Si$^+$ is at least 10$\times$ more populated than the excited
state.  While one could construct a scenario which reproduces the
observations for either excitation mechanism, we interpret the
overabundance of the ground level as evidence for UV pumping for two
reasons.  First, the effect is naturally explained within the context
of UV pumping by optical depth effects.  If the cloud is optically
thick, then one expects partial excitation especially if one cloud
`shadows' the rest of the gas that is further away from the afterglow.
Second, the similarity in kinematics suggest the gas is co-spatial and
yet under collisional excitation the excited gas would require far
greater pressure than is conceivable for the unexcited phase.

%Third, it would
%be unusual for the higher density phase to exhibit signficantly lower
%column density than the unexcited, lower density gas.

\begin{figure}[ht]
\begin{center}
\includegraphics[height=3.6in,angle=90]{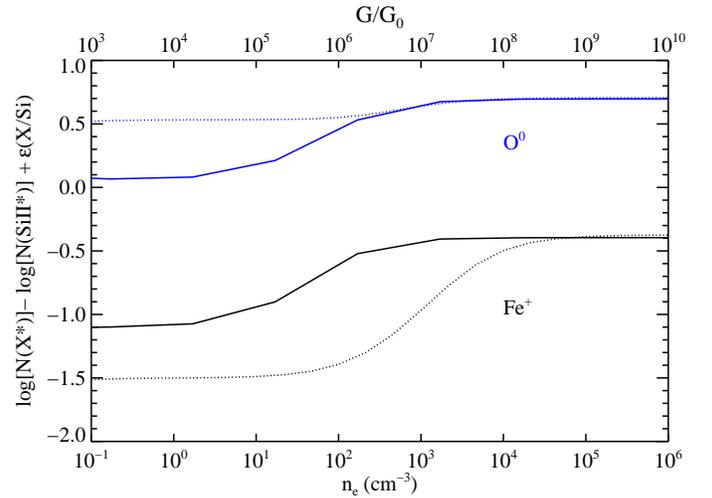}
\caption{Comparison of the predicted column densities of the
first excited states of 
Fe$^+$ (black) and O$^0$ (blue) relative to the $J=3/2$ excited
level of Si$^+$ for 
indirect UV pumping (solid curves) and collisional excitation (dotted; $T=2600$K).
We have assumed gas-phase abundances
$\epsilon({\rm O/Si}) = 1$ and $\epsilon({\rm Fe/Si}) = 0$.  
It is important to note that the
collisional excitation calculations 
for O$^0$ 
assume an ionization fraction $x=10^{-2}$.  Higher values of $x$
would lead to lower $\N{OI^*}$ values relative to $\N{SiII^*}$
in a roughly linear fashion.
}
\label{fig:disting}
\end{center}
\end{figure}

\subsection{When Excited States of Fe$^+$ are not Detected}

For GRB afterglows where the excited \ion{Fe}{2} levels are not detected, the
abundances of the first excited levels of O$^0$ and Si$^+$, together
with an upper limit in excited Fe$^+$ can be applied to constrain the
excitation mechanism.  Figure~\ref{fig:disting} presents the values of
$\noij{1}$ and $\nfej{7/2}$ relative to $\nsij{3/2}$ as a function of
electron density and intensity for collisional excitation (dotted
curves) and UV pumping (solid curves), respectively.  Note that the
curves assume that the gas-phase O$^0$ abundance is 10 times higher than
Si$^+$, $\epsilon({\rm O/Si}) \equiv \log\N{O^0} - \log\N{Si^+} = 1$,
and that Si$^+$ and Fe$^+$ have equal gas-phase abundance,
$\epsilon({\rm Fe/Si}) = 0$.  Differential depletion, in particular,
would modify these assumptions: increasing $\epsilon({\rm O/Si})$ and
decreasing $\epsilon({\rm Fe/Si})$.  
%At large density and intensity
%where the fine-structure levels saturate, the predictions are
%identical.  At lower values, the ratios of the Fe$^+$ and O$^0$
%excited states decline relative to Si$^+$ because the excitation rates
%are highest for Si$^+$
%(Figures~\ref{fig:uvpump},\ref{fig:collisions}).  Furthermore, we note
%that there are unique predictions for collisions and pumping.

We first focus on Fe$^+$ and Si$^+$.  It is evident that collisional
excitation at $n_e < 10^3$ allows for significantly lower values of
$\nfej{7/2}/\nsij{3/2}$ than UV pumping with $G/G_0\le 10^7$.  The
main result is that UV pumping does not allow $\log\nfej{7/2} -
\log\nsij{3/2} < [-1 + \epsilon({\rm Fe/Si})]$.  Therefore, a GRB
afterglow spectrum exhibiting \ion{Si}{2}* absorption and strict upper
limits on the excited \ion{Fe}{2} transitions is likely to indicate
collisional excitation.  There are, however, two caveats.  First, it
assumes knowledge of the inherent Fe/Si ratio. In general, this may be
inferred from the ground state populations.  Second, the results in
Figure~\ref{fig:disting} also assume steady-state conditions.  The
observations of GRB afterglows may occur on time-scales that are short
compared to the lifetime $\tau_{\rm life}$ of the excited levels
(especially for Si$^+$ and O$^0$ which have $\tau_{\rm life} = 1.3$
and 3.1 hr, respectively).  Therefore, a significant underabundance of
$\nfej{7/2}$ relative to $\nsij{3/2}$ is possible from UV pumping if
one observes the excited gas at a time when the excited
Fe$^+$ ions ($\tau_{\rm life} \approx 8$min) have decayed to the
ground-state.
In this case, of course,
one would expect line-strength variability between early- and late-time
spectroscopic data.

%We note, in passing, that the afterglow spectrum of GRB~050820
%offers a case study.  We detect $\nsij{3/2} = 10^{13.8} \cm{-2}$
%and $\nfej{7/2} < 10^{13.05} \cm{-2}$ (Prochaska et al.\ 2006, in prep).
%Unfortunately, the difference is consistent with both
%excitation mechanisms.

Next, we consider the O$^0$ and Si$^+$ predictions.  Independent of
the excitation mechanism, we find $\noij{1} \geq \nsij{3/2}$ for the
assumed ionization fraction\footnote{This is a rough estimate of the
ionization fraction for a predominantly neutral gas.}, 
$x \approx n_e/n_H = 10^{-2}$.  For a
predominantly neutral gas, collisional excitation of Si$^+$ will
roughly match that of O$^0$ and $\noij{1} > \nsij{3/2}$ is predicted.
For larger values of $x$, however, the excitation rate for Si$^+$ is
considerably higher and a significantly smaller ratio \\
of $\noij{1} / \nsij{3/2}$ is expected.  
In fact, the only process which allows
$\noij{1} < \nsij{3/2}$ is collisional excitation in a predominantly
ionized medium, e.g.\ the Carina nebula in the Galaxy
\citep{walborn02}.

%Finally, we note that the arguments made in the previous sub-section
%regarding atomic gas also apply here.
%Specifically, the detection of \ion{Si}{2}* transitions will
%likely imply $n_e > 10^2 \cm{-3}$ under the assumption of collisional
%excitation.  With this density, one may expect to observe 
%Mg$^0$, C$^0$, and possibly Fe$^0$ gas.

\section{DISCUSSION}
\label{sec:discuss}

%  Furthermore, the \ion{Si}{2}~1808 and
%\ion{Zn}{2} profiles found in GRB\,051111 are stronger than any known
%quasar absorption line system to date \citep{pro03}.  Finally while
%the strongest transitions (e.g.\ \ion{Mg}{2}~2796, \ion{Fe}{2}~2600)
%exhibit significant absorption over a velocity interval $\Delta
%v_{sat} \approx 200 \mkms$, the unsaturated profiles show that $>
%90\%$ of the gas is confined to a velocity width $\Delta v_{90} = 40
%\mkms$.

\subsection{Implications for GRB~051111 and GRB~050730}

Despite different spectral coverage in the rest frame of GRB\,051111
and GRB\,050730, the echelle spectra of these afterglows exhibit
ground-state Si$^+$ and Fe$^+$ features from all of their excited
levels.  In the case of GRB\,050730, excited O$^0$ is also present.
But due to measurement uncertainties, their population ratios may be
explained by Boltzmann distribution with $T_{Ex} = 2600 - 6000$K
(Figures 10 \& 11), as well as by UV pumping provided $G/G_0 \gtrsim
10^{6.5}$.  We have investigated variations in the column densities
and set {\it upper limits} to the line-strength variability of $< 10\%$ per
1800s for GRB~051111 and $< 30\%$ per 1800s for GRB~050730.
It appears difficult to unambiguously rule out either UV pumping or
collisional excitation as the dominant excitation mechanism.
The only strong argument against collisional excitation is the
absence of Fe$^0$ and C$^0$ atomic gas for GRB~051111 and
GRB~050730, respectively. 

Under collisional excitation, the excited ions occur irrespective of
the presence of the afterglow and the gas is required to have $n_e >
10^{5} \cm{-3}$ and $T \sim 2000-6000$ K to populate the \ion{Fe}{2}
excited levels.  If the excited gas arises from a layer of high
density ($n_e > 10^5 \cm{-3}$), then based on the absence of
\ion{Fe}{1} we can place an upper limit on the ratio of the electron
density to far-UV intensity $n_e\times (G/G_0)^{-1}$
(Equation~\ref{eqn:nefe}).  Adopting the best-fit Boltzmann
distribution, we calculate the total Fe$^+$ column density $N_{\rm
tot}{Fe^+} = 10^{14.4} \cm{-2}$, including all levels.  Adopting a
conservative upper limit to the atomic state $\N{Fe^0} < 10^{12}
\cm{-2}$ (Table~\ref{tab:051111}) which accounts for continuum and
statistical uncertainty, we therefore find from Equation~\ref{eqn:nefe}
that $n_e\times (G/G_0)^{-1} < 1$ for $T < 10000$K.  For $n_e > 10^5
\cm{-3}$, we find that $G/G_0 > 10^5$.  Therefore,
we conclude that we would only observe significant populations of the
excited \ion{Fe}{2} levels for gas in the presence of a strong far-UV
radiation field.  

To satisfy $G/G_0 > 10^5$, it requires the gas to lie within $\approx
0.1$pc of a $T=40000$K O star with $R = 10 R_\odot$.  While the GRB
progenitor star is expected to reside in a star-forming region and
therefore near other massive stars, we consider it highly unrealistic
that the gas is at $>100$ pc from the GRB but also within $0.1$ pc of
an O star.

A similar conclusion can be drawn for GRB~050730 using \ion{C}{1}.  We
infer a C$^+$ column density by scaling $\N{Fe+} = 10^{14.8} \cm{-2}$
from the excited region with solar relative abundances: $\N{C^+} =
10^{15.8} \cm{-2}$.  The observed upper limit $\N{C^0}/\N{C^+} <
10^{-2.3}$ requires $G/G_0 > 5 \sci{4}$, again assuming a conservative
limit to the temperature ($T < 10000$K).  For GRB~050730, we have no
significant constraint on the distance of the gas from the afterglow.
But, if collisional excitation dominates indirect UV pumping (and that
$n_e < 10^{9} \cm{-3}$), then this gas must also be experiencing a
strong local source of far-UV radiation.  We conclude based on the
null detections of atomic gas that {\it indirect UV pumping by GRB
afterglow radiation is the dominant excitation process.}

We also note that there are additional challenges to a collisional
excitation scenario.  First, the gas would require a very large
heating source to maintain the inferred temperatures for collisional
excitation.  For the implied physical conditions (e.g.\ $T=2600$ K and
$n_e > 10^5 \cm{-3}$), [\ion{Fe}{2}] cooling dominates.  Using the
coefficients given by \cite{hollenbach89}, we calculate a rate of $n
\lfeII = 4 \sci{-23} {\rm ergs \, s^{-1} \, H^{-1}}$ assuming 1/10
solar metallicity and that $90\%$ of the iron is locked into dust
grains.  This implies a cooling rate $t_{cool} \sim kT / (n \lfeII) <
500 {\rm yr^{-1}}$.  Including the cooling by meta-stable states of
\ion{O}{1} would imply an even shorter cooling time.  Second, gas with
$T \approx 3000$ K is generally not stable to thermal perturbation.
In the case considered here, the [\ion{Fe}{2}] cooling curve is
relatively insensitive to temperature variations but very sensitive to
density changes.  For example, an increase in $n_H$ increases $n
\lfeII$ and drives the gas to higher density and lower temperature
until $T \ll 1000$\,K.  Third, the implied pressure $nT$ is extremely
high.  One would require a strong mechanical process (e.g.\ shock) to
confine the gas and it would be surprising to therefore find the
quiescent kinematic characteristics that are observed
(Figures~\ref{fig:051111},\ref{fig:050730}).

Adopting UV pumping as the the excitation mechanism, we can constrain
the distance of the gas from the GRB based on known afterglow
radiation field.  For GRB~051111, the gas must arise at $r \approx
200$\,pc.  This distance is large enough to prevent the ionization of
Mg$^0$ but also close enough that the \ion{Fe}{2} and \ion{Si}{2} levels would
be UV pumped and also show the relative populations that are observed.
In the case of GRB~050730, where the excited \ion{Fe}{2} levels are
populated according to their degeneracies $g_J$, the gas must lie
within $\approx 100$\,pc of the afterglow (Table~\ref{tab:rlim}).  The
ionizing column of photons at 1\,pc is $6\sci{22} \cm{-2}$.  The gas
is therefore not likely at $< 3$\,pc, where the far-UV radiation field
from the GRB afterglow will photoionize all of the C$^+$, Si$^+$,
O$^0$, and Fe$^+$ gas.
%Therefore, while UV pumping
%shrouds the physical conditions of the fine-structure gas (i.e.\ $n_e,
%T_{Ex}$), it provides additional constraints on the distance from the
%GRB afterglow.  %[anything else?]

\subsection{Implications of Fine-Structure Lines in Other GRB
Afterglows}

The detection of \ion{Si}{2}* transitions is nearly ubiquitous
in GRB afterglow spectra \citep{vel+04,cpb+05,bpck+05}.
This stands in stark contrast with intervening quasar absorption
line systems where only \ion{C}{1}* and \ion{C}{2}* fine-structure
lines have been observed \citep[e.g.][]{srianand00,wolfe03a,howk05}.
To date, the observation of \ion{Si}{2}* absorption for GRB have
been interpreted as the result of collisional excitation and therefore
evidence for large volume densities in the gas.
We have demonstrated in this paper, however, that UV pumping is the
most likely excitation mechanism if the gas is located within
a distance of $\approx 100$\,pc.  This conclusion holds for all
of the GRBs with reported \ion{Si}{2}* detections
(Table~\ref{tab:rlim}) except for GRB~050408 \citep{foley06}.
And, for sightlines exhibiting \ion{Fe}{2}* absorption, we
contend UV pumping is likely the only mechanism.

In summary, we conclude that in the presence of an intensified UV
radiation field from the afterglow {\it indirect UV pumping alone
offers a simple explanation for all the excited fine-structure
transitions observed in GRB host environment to date without invoking
extreme ISM properties}, such as high gas density.  If we take the gas
density allowed by the observed relative abundances of atomic species,
$n_{\rm H}=100-1000\,cm^{-3}$ (\S\ 4.2), the observed large \nhi\
\citep[e.g.][]{cpb+05} implies a cloud thickness of $3-30$ pc which is
typical of giant molecular clouds (Turner 1988).  In lieu of
diagnostics from the fine-structure populations, one must rely on
comparisons of the atomic states with higher ionization levels to
constrain the gas density.  We contend that the principal quantity
inferred from the fine-structure lines is the distance to the GRB
afterglow.  At present, there is no obvious diagnostic for the gas
temperature.

\subsection{Implications for Gas Within 100\,pc of GRB Afterglows}

Our analysis reveals a number of implications for the study of gas
located within 100\,pc of the GRB afterglow.  The first is that this
gas may be identified by the absence of atomic absorption, in
particular \ion{Mg}{1} gas.  Although the sightline is likely to
penetrate \ion{Mg}{1} gas at distances far from the GRB (e.g.\ within
the galactic halo), differences in the line-of-sight velocities should
distinguish these clouds.  Therefore, a study of the kinematic
differences between \ion{Mg}{1} and \ion{Mg}{2} profiles in GRB
spectra is warranted.

In the absence of atomic gas (e.g.\ \ion{Mg}{1} absorption) that
coincides kinematically with the low-ion species, the gas may have
small distance from the GRB afterglow and therefore may be
circumstellar material from the progenitor.  Unfortunately, the
afterglow radiation washes out the most readily available diagnostics
for studying the density and temperature of this gas.  We note,
however, that self-shielding of UV photons matching various excitation
energies may be significant, especially for the majority of other GRB
sightlines where the fine-structure column density is signficantly
lower than the ground-state \citep[e.g.][]{vel+04,savaglio04}.  We
expect this may also be the case for GRB~051111 and GRB~050730.  It is
therefore possible that a significant portion of the gas is not
excited by the afterglow and we could infer physical conditions for
the bulk of the medium.

Another implication is that one must consider the fine-structure
levels in the chemical abundance analysis.  This will be especially
important for sightlines with low total metal-line column density.
The correction to $\N{Fe^+}$ in GRB~050730, for example, is 33$\%$ and
we estimate the correction to $\N{Si^+}$ is greater than 50$\%$.  
If one cannot account for the excited-states in the analysis (e.g.\ 
because the lines are too saturated/blended for accurate measurement), 
then analysis of
S$^+$ is preferred because it does not have fine-structure levels
near the ground-state.

Finally, we emphasize again that to observe gas at 10\,pc one requires
a large pre-existing column density of neutral or partially ionized
gas prior to the GRB event.  This includes highly ionized species like
\ion{C}{4}.  Consider, as an example, the relatively spectacular
\ion{C}{4} gas observed toward GRB~021004 which several authors have
interpreted to arise from gas associated with the GRB progenitor
\citep{sgh+03,Mir03,vlg05,fdl+05}.  At $t_{obs}=1$ hr after the GRB,
we estimate an \ion{H}{1} ionizing column density of $3 \sci{21} \,
{\rm photons} \, \cm{-2}$ at 10\,pc assuming the afterglow properties
given in Table~\ref{tab:rlim}.  This column significantly exceeds the
\nhi\ value observed for the gas toward GRB~021004, therefore we
assume that all of the low-ion gas within 10\,pc was either ionized
prior to the GRB or prior to the observations.  This helps explain the
absence of strong low-ion absorption\footnote{The observation of
\ion{Al}{2} absorption is an interesting exception.}.  More
importantly, the column density of C$^{+3}$ ionizing photons with
energy $h\nu = 64.5$ to 70eV at 10\,pc is $\approx 10^{20} \cm{-2}$.
If this flux is unattenuated by gas within 10\,pc then it would
significantly ionize all of the C$^{+3}$ ions within a few tens of pc
from the GRB.  At these energies, the principal source of opacity is
He$^+$, but we estimate that this gas will also have been burned away
by the GRB and its progenitor.  We contend, therefore, that the
majority of C$^{+3}$ gas observed along the GRB~021004 sightline is
either associated with the larger star forming region and/or halo gas
from the GRB host galaxy and/or its galactic neighbors.

\acknowledgments

The authors wish to recognize and acknowledge the very significant
cultural role and reverence that the summit of Mauna Kea has always
had within the indigenous Hawaiian community.  We are most fortunate
to have the opportunity to conduct observations from this mountain.
We acknowledge the comments of an anonymous referee and also
D. Hollenbach who stressed the importance of UV pumping.  We are
grateful to Grant Hill, Derek Fox and Barbara Schaefer for their roles
in obtaining the Keck/HIRES data of GRB\,050111 and I. Thompson for
obtaining the Magellan/MIKE data of GRB\,050730.  The authors would
like to thank C. McKee, N. Walborn, T. Gull, K. Sembach, A. Wolfe,
B. Mathews, D. York, D. Welty, C. Howk, A. Konigl, and
G. Blumenthal for helpful discussions.  %We give thanks to
%N. Walborn and T. Gull for providing HST/STIS ultraviolet spectra of
%the carnia nebula and $\eta$ carina.  
We thank Weidong Li, Nat
Butler, and Dan Perley for providing preliminary results on 
the light curves of GRB~0501111 and GRB~060206.  
We thank Miroslava Dessauges-Zavadsky for providing UVES
measurements.  
J.X.P., H.-W.C., and J.S.B. are partially supported by \\
NASA/Swift grant NNG05GF55G.

\appendix

\section{Photoionization Equilibrium}
\label{app:atomic}

Under steady-state conditions and photoionization equilibrium, the balance of  
photoionization and recombination for Mg$^0$ and Mg$^+$ gives

\begin{equation}
n({\rm Mg^0}) n_\gamma\,c\,\sigma_{ph}^{\rm Mg^0} =
n({\rm Mg^+}) n_e \ltk \alpha_{r}(T) + \alpha_{di}(T) \rtk 
\label{eqn:mgbal}
\end{equation}

\noindent where $\alpha_r$ and $\alpha_{di}$ are the radiative
and dielectronic recombination coefficients and $\sigma_{ph}^{\rm Mg^0}$ is
the cross-section to photoionization integrated over the 
incident radiation field.
For the recombination rates, we will use the fitting formula given
by \cite{aldrovandi73}, 

\begin{equation}
\alpha_r^{\rm Mg^+}(T) = 1.4 \sci{-13} (T/10^4)^{-0.855} {\rm cm^3 \, s^{-1}} \cmma
\end{equation}

\noindent
and \cite{shull82}, 

\begin{equation}
\alpha_{di}^{\rm Mg^+}(T) = 4.49\sci{-4} T^{-1.5} \exp(-5.01\sci{4}/T) 
\ltk 1 + 0.0021 \exp(-2.61\sci{4} /T) \rtk {\rm cm^3 \, s^{-1}} \perd
\end{equation}

\noindent For the photoionization rate, we integrate the Galactic
far-UV radiation field determined by \cite{gpw80} over the fitting function
for $\sigma_{ph}^{\rm Mg^0}$ provided by \cite{verner96}:

\begin{equation}
n_\gamma c \sigma_{ph}^{\rm Mg^0} = 
\Gamma({\rm Mg^0}) = 6.8 \sci{-11} \frac{G}{G_0} 
\; {\rm s^{-1}} \perd
\label{eqn:ionmg}
\end{equation}

\noindent Although the radiation field for the gas associated with
a GRB host galaxy will likely have a different spectral index than that
of the Milky Way, the ionization rate is 
dominated by the photon density at $h \nu \approx 8$\,eV and, in any
case, 
uncertainties in the atomic data limit the accuracy of this 
analysis to factors of a few.
Rearranging Equations~\ref{eqn:mgbal}-\ref{eqn:ionmg} to
express $n_e$ in terms of the observed $\N{Mg^0}/\N{Mg^+}$ ratio
we reproduce Equation~\ref{eqn:nemg}.

One can derive a similar expression for Fe.  In this case, radiation
recombination dominates for $T < 8000$K, 

\begin{equation}
\alpha_r^{\rm Fe^+}(T) = 8.945 \sci{-9} \ltk (\sqrt{T/0.04184} 
(\sqrt{T/0.04184} +1)^{0.7844} (1 + \sqrt{T/5.35\sci{13}})^{1.2156} \rtk^{-1}
\end{equation}
For the ionization rate we perform the same calculation as above 
(Equation~\ref{eqn:ionmg}) and find

\begin{equation}
\Gamma({\rm Fe^+}) = 1.7 \sci{-10} (G/G_0) \perd
\end{equation}

Finally, here are the relevant expressions for Carbon:

\begin{equation}
\alpha_r^{\rm C^+}(T) = 7.651 \sci{-9} \ltk (\sqrt{T/0.001193} 
(\sqrt{T/0.001193} +1)^{0.1983} 
(1 + \sqrt{T/9.334\sci{12}})^{1.8027} \rtk^{-1}
\end{equation}

\noindent and

\begin{equation}
\Gamma({\rm C^+}) = 6.6 \sci{-10} (G/G_0) \perd
\end{equation}

\clearpage

%\bibliographystyle{../apj}
%\bibliography{../journals_apj,../../grbrefs,../qal,../ism}

\begin{thebibliography}{}

\bibitem[{Aldrovandi} \& {Pequignot}(1973)]{aldrovandi73}
{Aldrovandi}, S.~M.~V. and {Pequignot}, D. 1973, \aap, 25, 137.

\bibitem[{Bahcall} \& {Wolf}(1968)]{bw68}
{Bahcall}, J.~N. and {Wolf}, R.~A. 1968, \apj, 152, 701.

\bibitem[{Barth} {\it et al.}\ (2003)]{bsc+03}
{Barth}, A.~J. {\it et al.}\  2003, ApJ ({\it Letters}), 584, L47.

\bibitem[{Berger} {\it et al.}\ (2005)]{bpck+05}
{Berger}, E. {\it et al.}\  2005, ApJ, in press (astro-ph/0511498)

\bibitem[{Bergeson} \& {Lawler}(1993)]{bergeson93}
{Bergeson}, S.~D. and {Lawler}, J.~E. 1993, \apjl, 414, L137.

\bibitem[{Bergeson}, {Mullman} \& {Lawler}(1994)]{bml94}
{Bergeson}, S.~D., {Mullman}, K.~L., and {Lawler}, J.~E. 1994, \apjl, 435,
  L157.

\bibitem[{Bergeson}, {Mullman} \& {Lawler}(1996)]{bergeson96}
{Bergeson}, S.~D., {Mullman}, K.~L., and {Lawler}, J.~E. 1996, \apj, 464, 1050.

\bibitem[{Bernstein} {\it et al.}\ (2003)]{bernstein03}
{Bernstein}, R. {\it et al.}\  2003, in { Instrument Design and Performance for
  Optical/Infrared Ground-based Telescopes. Edited by Iye, Masanori; Moorwood,
  Alan F. M. Proceedings of the SPIE, Volume 4841, pp. 1694-1704 (2003).},
  1694.

\bibitem[{Bloom} {\it et al.}\ (2002)]{bkp+02}
{Bloom}, J.~S. {\it et al.}\  2002, ApJ ({\it Letters}), 572, L45.

\bibitem[{Butler} {\it et al.}\ (2006)]{butler06}
{Butler}, N. {\it et al.}\  2006, In prep.

\bibitem[{Cenko} \& {Fox}(2005)]{gcn3834}
{Cenko}, S.~B. and {Fox}, D.~B. 2005, GRB Coordinates Network, 3834, 1.

\bibitem[{Charlton} \& {Churchill}(1998)]{charlton98}
{Charlton}, J.~C. and {Churchill}, C.~W. 1998, \apj, 499, 181.

\bibitem[{Chen} {\it et al.}\ (2005)]{cpb+05}
{Chen}, H.-W. {\it et al.}\  2005, \apjl, 634, L25.

\bibitem[{Christensen}, {Hjorth} \& {Gorosabel}(2004)]{chg04}
{Christensen}, L., {Hjorth}, J., and {Gorosabel}, J. 2004, A\&A, 425, 913.

\bibitem[{Churchill} {\it et al.}\ (2000)]{churchill00a}
{Churchill}, C.~W. {\it et al.}\  2000, \apjs, 130, 91.

\bibitem[{Draine} \& {Hao}(2002)]{draine02}
{Draine}, B.~T. and {Hao}, L. 2002, \apj, 569, 780.

\bibitem[{Fiore} {\it et al.}\ (2005)]{fdl+05}
{Fiore}, F. {\it et al.}\  2005, \apj, 624, 853.

\bibitem[{Foley} {\it et al.}\ (2005)]{foley06}
{Foley}, R.~J. {\it et al.}\  2006, ApJ, in press (astro-ph/0512081) 

\bibitem[{Frisch} {\it et al.}\ (1990)]{frisch90}
{Frisch}, P.~C. {\it et al.}\  1990, \apj, 357, 514.

\bibitem[{Frisch}, {York} \& {Fowler}(1987)]{frisch87}
{Frisch}, P.~C., {York}, D.~G., and {Fowler}, J.~R. 1987, \apj, 320, 842.

\bibitem[{Gondhalekar}, {Phillips} \& {Wilson}(1980)]{gpw80}
{Gondhalekar}, P.~M., {Phillips}, A.~P., and {Wilson}, R. 1980, \aap, 85, 272.

\bibitem[{Gull} {\it et al.}\ (2005)]{gull05}
{Gull}, T.~R. {\it et al.}\  2005, \apj, 620, 442.

\bibitem[{Habing}(1968)]{habing68}
{Habing}, H.~J. 1968, \bain, 19, 421.

\bibitem[{Hall} {\it et al.}\ (2002)]{has+02}
{Hall}, P.~B. {\it et al.}\  2002, \apjs, 141, 267.

\bibitem[{Hill} {\it et al.}\ (2005)]{gcn4255}
{Hill}, G. {\it et al.}\  2005, GRB Circular Network, 4255, 1.

\bibitem[{Hollenbach} \& {McKee}(1989)]{hollenbach89}
{Hollenbach}, D. and {McKee}, C.~F. 1989, \apj, 342, 306.

\bibitem[{Howk}, {Wolfe} \& {Prochaska}(2005)]{howk05}
{Howk}, J.~C., {Wolfe}, A.~M., and {Prochaska}, J.~X. 2005, \apjl, 622, L81.

\bibitem[{Jenkins}(1996)]{jenkins96}
{Jenkins}, E.~B. 1996, \apj, 471, 292.

\bibitem[{Lagrange-Henri}, {Vidal-Madjar} \& {Ferlet}(1988)]{lvf88}
{Lagrange-Henri}, A.~M., {Vidal-Madjar}, A., and {Ferlet}, R. 1988, \aap, 190,
  275.

\bibitem[{Le Floc'h} {\it et al.}\ (2006)]{lefloch06}
{Le Floc'h}, E. {\it et al.}\  2006, (astro-ph/0601252)

\bibitem[{Lipkin} {\it et al.}\ (2004)]{log+04}
{Lipkin}, Y.~M. {\it et al.}\  2004, \apj, 606, 381.

\bibitem[{Mirabal} {\it et al.}\ (2003)]{Mir03}
{Mirabal}, N. {\it et al.}\  2003, \apj, 595, 935.

\bibitem[{Mirabal} {\it et al.}\ (2002)]{mhk+02}
{Mirabal}, N. {\it et al.}\  2002, ApJ, 578, 818.

\bibitem[{Morton}(1991)]{morton91}
{Morton}, D.~C. 1991, \apjs, 77, 119.

\bibitem[{Morton}(2003)]{morton03}
{Morton}, D.~C. 2003, \apjs, 149, 205.

\bibitem[{Nielsen}, {Gull} \& {Vieira Kober}(2005)]{nielsen05}
{Nielsen}, K.~E., {Gull}, T.~R., and {Vieira Kober}, G. 2005, \apjs, 157, 138.

\bibitem[Paczy\'nski(1998)]{pac98a}
Paczy\'nski 1998, in { Gamma Ray Bursts: 4th Huntsville Symposium}, volume 428,
  (Woodbury, New York: AIP), 783.

\bibitem[{Penprase} {\it et al.}\ (2005)]{penprase05}
{Penprase}, B.~E. {\it et al.}\  2005, ApJ, in press (astro-ph/0512340)

\bibitem[{Perley} {\it et al.}\ (2006)]{danp06}
{Perley}, N. {\it et al.}\  2006, In prep.

\bibitem[{Perna} \& {Loeb}(1998)]{pl98a}
{Perna}, R. and {Loeb}, A. 1998, ApJ, 501, 467.

\bibitem[{Prochaska}(2005)]{gcn4271}
{Prochaska}, J. 2005, GRB Circular Network, 4271, 1.

\bibitem[{Prochaska} \& {Wolfe}(2002)]{pw02}
{Prochaska}, J.~X. and {Wolfe}, A.~M. 2002, \apj, 566, 68.

\bibitem[{Raassen} \& {Uylings}(1998)]{raassen98}
{Raassen}, A.~J.~J. and {Uylings}, P.~H.~M. 1998, \aap, 340, 300.

\bibitem[{Ramirez-Ruiz} {\it et al.}\ (2005)]{rgs+05}
{Ramirez-Ruiz}, E. {\it et al.}\  2005, \apj, 631, 435.

\bibitem[{Rao} \& {Turnshek}(2000)]{rao00}
{Rao}, S.~M. and {Turnshek}, D.~A. 2000, \apjs, 130, 1.

\bibitem[{Sarazin}, {Rybicki} \& {Flannery}(1979)]{sarazin79}
{Sarazin}, C.~L., {Rybicki}, G.~B., and {Flannery}, B.~P. 1979, \apj, 230, 456.

\bibitem[{Savage} \& {Sembach}(1991)]{savage91}
{Savage}, B.~D. and {Sembach}, K.~R. 1991, \apj, 379, 245.

\bibitem[{Savaglio} \& {Fall}(2004)]{savaglio04}
{Savaglio}, S. and {Fall}, S.~M. 2004, \apj, 614, 293.

\bibitem[{Savaglio}, {Fall} \& {Fiore}(2003)]{sff03}
{Savaglio}, S., {Fall}, S.~M., and {Fiore}, F. 2003, ApJ, 585, 638.

\bibitem[{Schaefer} {\it et al.}\ (2003)]{sgh+03}
{Schaefer}, B.~E. {\it et al.}\  2003, \apj, 588, 387.

\bibitem[{Shull} \& {van Steenberg}(1982)]{shull82}
{Shull}, J.~M. and {van Steenberg}, M. 1982, \apjs, 48, 95.

\bibitem[{Silva} \& {Viegas}(2001)]{silva01}
{Silva}, A.~I. and {Viegas}, S.~M. 2001, Computer Physics Communications, 136,
  319.

\bibitem[{Silva} \& {Viegas}(2002)]{silva02}
{Silva}, A.~I. and {Viegas}, S.~M. 2002, \mnras, 329, 135.

\bibitem[{Srianand}, {Petitjean} \& {Ledoux}(2000)]{srianand00}
{Srianand}, R., {Petitjean}, P., and {Ledoux}, C. 2000, \nat, 408, 931.

\bibitem[{Stanek} {\it et al.}\ (2003)]{smg+03}
{Stanek}, K.~Z. {\it et al.}\  2003, ApJ ({\it Letters}), 591, L17.

\bibitem[{Steidel} \& {Sargent}(1992)]{steidel92}
{Steidel}, C.~C. and {Sargent}, W.~L.~W. 1992, \apjs, 80, 1.

\bibitem[{Tripp}, {Lu} \& {Savage}(1996)]{tripp96}
{Tripp}, T.~M., {Lu}, L., and {Savage}, B.~D. 1996, \apjs, 102, 239.

\bibitem[{van Marle}, {Langer} \& {Garcia-Segura}(2005)]{vlg05}
{van Marle}, A.-J., {Langer}, N., and {Garcia-Segura}, G. 2005,
\aap, 444, 837.


\bibitem[{Verner} {\it et al.}\ (1996)]{verner96}
{Verner}, D.~A. {\it et al.}\  1996, \apj, 465, 487.

\bibitem[{Vink}(2005)]{vink05}
{Vink}, J.~S. 2005, (astro-ph/0511048)

\bibitem[{Vogt} {\it et al.}\ (1994)]{vogt94}
{Vogt}, S.~S. {\it et al.}\  1994, in { Proc. SPIE Instrumentation in Astronomy
  VIII, David L. Crawford; Eric R. Craine; Eds., Volume 2198, p. 362}, 362.

\bibitem[{Vreeswijk} {\it et al.}\ (2004)]{vel+04}
{Vreeswijk}, P.~M. {\it et al.}\  2004, \aap, 419, 927.

\bibitem[{Walborn} {\it et al.}\ (2002)]{walborn02}
{Walborn}, N.~R. {\it et al.}\  2002, \apjs, 140, 407.

\bibitem[{Welty}, {Hobbs} \& {Morton}(2003)]{welty03}
{Welty}, D.~E., {Hobbs}, L.~M., and {Morton}, D.~C. 2003, \apjs, 147, 61.

\bibitem[{Wijers}(2001)]{wij01}
{Wijers}, R.~A.~M.~J. 2001, in { Gamma-Ray Bursts in the Afterglow Era,
  Proceedings of the International workshop held in Rome, CNR headquarters,
  17--20 October, 2000}, ed.\ Enrico Costa, Filippo Frontera, and Jens Hjorth,
  (Berlin Heidelberg: Springer), 306.

\bibitem[{Wolfe}, {Prochaska} \& {Gawiser}(2003)]{wolfe03a}
{Wolfe}, A.~M., {Prochaska}, J.~X., and {Gawiser}, E. 2003, \apj, 593, 215.

\bibitem[Woosley(1993)]{woo93}
Woosley, S.~E. 1993, ApJ, 405, 273.

\bibitem[{York} \& {Kinahan}(1979)]{york79}
{York}, D.~G. and {Kinahan}, B.~F. 1979, \apj, 228, 127.

\end{thebibliography}

\end{document}